\documentclass[preprint,11pt,authoryear,sort&compress]{elsarticle}
\journal{Journal of the Mechanics and Physics of Solids}
\usepackage{amssymb,amsmath}
\usepackage{hyperref}
\usepackage{caption}
\usepackage{float}
\usepackage{booktabs}
\usepackage{longtable}
\usepackage{indentfirst} 
\usepackage{geometry}
\usepackage{enumitem}
\usepackage{hyperref}
\hypersetup{hidelinks}
\usepackage{xurl}
\usepackage[titletoc]{appendix}
\usepackage{graphicx}
\usepackage{latexsym}
\usepackage{subfigure}
\usepackage{wrapfig}

\usepackage{comment}
\usepackage{soul}
\usepackage{tablefootnote}
\usepackage{multirow}
\usepackage{adjustbox}

\newcommand{\bild}[6]       % Name, Beschriftung, Referenz, Breite, Platzierung
{
     \begin{figure}[#5]                         
      \begin{center}
        \includegraphics*[width=#4]{#1}
        \caption[#2]{\label{#3} #2 #6}
      \end{center}
     \end{figure}
}

\usepackage[usenames,dvipsnames]{xcolor}
\begin{document}
\captionsetup[figure]{labelfont={bf},labelformat={default},labelsep=period,name={Fig.}}
\begin{frontmatter}
\title{Coupling Diffusion and Finite Deformation in Phase Transformation Materials}

%% or include affiliations in footnotes:
\author[mymainaddress]{Tao Zhang}

 \ead{taozhang@ucsb.edu}

\author[mysecondaryaddress]{Delin Zhang}
 \ead{delinzha@usc.edu}

\author[mymainaddress]{Ananya Renuka Balakrishna\corref{cor1}}
\cortext[cor1]{Corresponding author.}
 \ead{ananyarb@ucsb.edu}

\address[mymainaddress]{Materials Department, University of California Santa Barbara, USA}
\address[mysecondaryaddress]{Department of Aerospace and Mechanical Engineering, University of Southern California, USA}
\begin{abstract} 
We present a multiscale theoretical framework to investigate the interplay between diffusion and finite lattice deformation in phase transformation materials. In this framework, we use the Cauchy-Born Rule and the Principle of Virtual Power to derive a thermodynamically consistent theory coupling the diffusion of a guest species (Cahn-Hilliard type) with the finite deformation of host lattices (nonlinear gradient elasticity). We adapt this theory to intercalation materials---specifically Li$_{1-2}$Mn$_2$O$_4$---to investigate the delicate interplay between Li-diffusion and the cubic-to-tetragonal deformation of lattices. Our computations reveal fundamental insights into the microstructural evolution pathways under dynamic discharge conditions, and provide quantitative insights into the nucleation and growth of twinned microstructures during intercalation. Additionally, our results identify regions of stress concentrations (e.g., at phase boundaries, particle surfaces) that arise from lattice misfit and accumulate in the electrode with repeated cycling. These findings suggest a potential mechanism for structural decay in Li$_2$Mn$_2$O$_4$. More generally, we establish a theoretical framework that can be used to investigate microstructural evolution pathways, across multiple length scales, in first-order phase transformation materials.
\end{abstract} 
\begin{keyword}
%% keywords here, in the form: keyword \sep keyword
First-order phase transformation \sep Lattice deformation \sep Intercalation material \sep Phase-field methods  \sep Energy storage
%% PACS codes here, in the form: \PACS code \sep code
%
%% MSC codes here, in the form: \MSC code \sep code
%% or \MSC[2008] code \sep code (2000 is the default)
%
\end{keyword}
\end{frontmatter}
\section{Introduction}

\noindent First-order phase transformation materials undergo an abrupt change in lattice geometries at critical temperature, stress, or composition values. In intercalation compounds, a type of first-order phase transformation material, lattices deform abruptly and often anisotropically when guest species (e.g., ions, atoms, or molecules) are inserted into the material \citep{padhi1997phospho, whittingham1978chemistry}. The reversible insertion of guest species makes intercalation materials suitable for applications in energy storage, optoelectronics, and catalysis \citep{wan2016tuning}, and are widely used as electrodes in rechargeable batteries. The reversible lattice deformation is commonly of two types: First, the deformation is dilatational (e.g., in LiFePO$_4$ or LiCoO$_2$ compounds) in which the unit cells expand/contract without a change in symmetry \citep{padhi1997phospho}. Second, the deformation is symmetry-lowering in which unit cells undergo a change in symmetry type (e.g., cubic-to-tetragonal in Li$_2$Mn$_2$O$_4$) during transformation \citep{thackeray1983lithium}. These lattice deformations are viewed as the root cause leading to the structural decay of intercalation materials and are often suppressed during reversible cycling \citep{bai2011suppression,zhang2021film}.

\vspace{2mm}
\noindent By contrast, we show that symmetry-lowering lattice deformations can be systematically designed to mitigate structural degradation in intercalation materials \cite{zhang2023designing, renuka2022crystallographic, schofield2022doping}. We developed algorithms to screen lattice deformations in $n > 5,000$ pairs of intercalation compounds and, found that several intercalation materials, such as the Spinels and NASICONs, undergo a symmetry-lowering transformation and can form shape-memory-like microstructures \cite{zhang2023designing}. These lattice deformations can be designed (e.g., through substitutional doping \cite{schofield2022doping, santos2023chemistry}), to minimize volume changes and interfacial stresses. However, to guide this design methodology, we need a robust and quantitative theoretical model that predicts how individual lattices deform and interact with the Li-diffusion front at the atomic scale, and how this interplay in turn governs the macroscopic material response. In this work, we develop a theoretical framework to investigate the interplay between diffusion and finite lattice deformation in symmetry-lowering phase transformation materials.

\begin{figure}[ht!]
    \centering
    \includegraphics[width=0.5\textwidth]{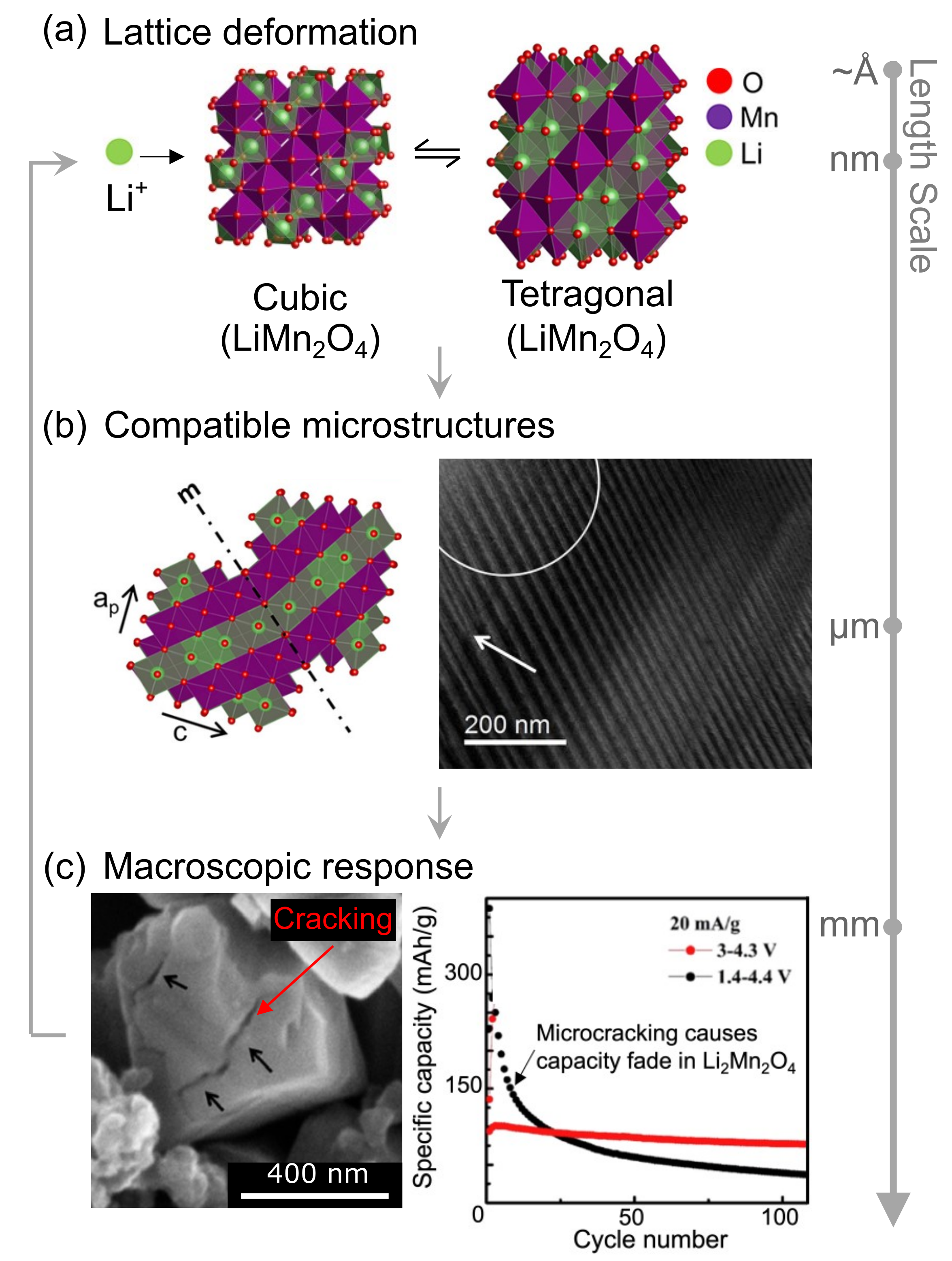}
    \caption{(a) Lithium intercalation into the host LiMn$_2$O$_4$ induces an abrupt Jahn-Teller deformation of the lattices at the atomic scale. This cubic-to-tetragonal deformation generates three lattice variants \citep{erichsen2020tracking}. (b) Individual lattices rotate and shear to form compatible interfaces called twin boundaries. At the mesoscale, this collective deformation of lattices in Li$_2$Mn$_2$O$_4$ generates a finely twinned pattern that resembles martensite-like microstructure in ferroelastic materials \citep{erichsen2020tracking} (Reprinted with permission from American Chemical Society). (c) The lattice misfit at the  LiMn$_2$O$_4$/ Li$_2$Mn$_2$O$_4$ phase boundary generates significant stresses which contribute to microcracking and eventual failure of intercalation compounds \citep{luo2020operando}(Reprinted with permission from Science Press and Dalian Institute of Chemical Physics, Chinese Academy of Sciences). In this work, we hypothesize that to conclusively understand the origins of structural degradation in intercalation materials, such as Li$_2$Mn$_2$O$_4$, we need to investigate the interplay between lattice deformations (atomic scale) and microstructural patterns (mesoscale) and how they collectively shape the macroscopic material response.}
    \label{Fig1}
\end{figure}

\vspace{2mm}
\noindent Symmetry-lowering phase transformations generate complex microstructural patterns, see Fig.~\ref{Fig1}(a-b) \citep{erichsen2020tracking}. The origins of these finely twinned domains can be explained as a consequence of energy minimization \citep{ball1987fine}. For example, the cubic to tetragonal transformation in Li$_{1-2}$Mn$_2$O$_4$ generates significant misfit strains between the cubic-LiMn$_2$O$_4$ and the tetragonal-Li$_2$Mn$_2$O$_4$ lattices. To minimize this elastic energy tetragonal lattices rotate and shear to form twinned domains at the continuum scale. This finely twinned mixture reduces the average misfit strains with the cubic phase, and thereby, minimizes the total elastic energy stored in the system. These microstructures can be analyzed using elastic energy arguments, however, it is important to understand how these microstructures interact with the diffusion of guest-species (e.g., Li-ions) and how this interplay, in turn, governs macroscopic material response such as internal stresses and micro-cracking, see Fig.~\ref{Fig1}(c) \citep{luo2020operando, renuka2022crystallographic}. 

\vspace{2mm}
\noindent At present, mesoscale models predict phase transformations in intercalation materials using guest-species composition (e.g., Li-ion) as the order parameter \citep{zhang2019phase, tang2010modeling, nadkarni2019modeling, ombrini2023thermodynamics, han2004electrochemical, zhang2020mechanically}. In these methods lattice deformations are typically homogenized and the free energy potential does not distinguish between the different lattice variants generated during symmetry-lowering phase transformations. Consequently, these models predict phase separation morphologies as a function of a scalar composition variable and are not suitable for investigating the rich heterogeneity in lattice deformations. These phase-field methods effectively predict reaction heterogeneities \citep{han2004electrochemical}, redox potentials \citep{ombrini2023thermodynamics}, and particle size effects \citep{tang2010modeling, zhang2019phase} under electrochemical operating conditions, however, they do not account for higher-order energy terms necessary to predict how twinned microstructures nucleate and grow during phase transformations. 

\vspace{2mm}
\noindent Researchers have developed phenomenological methods that couple strain and composition fields within a single framework \citep{rudraraju2014three, rudraraju2016mechanochemical, balakrishna2018combining, balakrishna2019phase}. For example, a chemo-mechanical model based on strain gradient elasticity theory describes the diffusion-driven martensitic phase transformations in multi-component crystalline solids \citep{rudraraju2016mechanochemical}.  In another example, we combined a Cahn-Hilliard model with a phase-field crystal model to investigate diffusion-induced stresses in binary alloys \citep{balakrishna2018combining}. These methods provide important insights into the coupling between higher-order diffusion and nonlinear strain gradient terms. These models, however, are based on variational derivations of the free energy functions that require a-priori specification and do not rigorously account for rate terms in deriving the governing equations \citep{gurtin1996generalized, gurtin2002gradient, di2014cahn,anand2012cahn}. Moreover, the dynamic electrochemical operating conditions driving phase transformations in intercalation compounds are not formulated in these frameworks. We will build on these efforts, to derive a thermodynamically consistent multiscale theory to investigate the diffusion-deformation interplay in phase transformation materials.

\vspace{2mm}
\noindent Our central aim is to establish a thermodynamically consistent framework that couples the diffusion of guest-species with the finite deformation of host lattices in phase transformation materials. To this end, we use the Cauchy-Born Rule and the Principle of Virtual Power to systematically derive the form of the free energy function and the governing equations for the diffusion-deformation model. We adapt this theoretical framework to intercalation materials by introducing specialized constitutive equations to capture the electrochemical operating conditions. We next solve this theoretical framework using a mixed-type finite element formulation based on Lagrange multipliers and calibrate the model to a spinel-type Li$_{2x}$Mn$_2$O$_4$ electrode as a representative example. Our results yield fundamental insights into the interplay between Li-diffusion and heterogeneous lattice deformations in Li$_{2x}$Mn$_2$O$_4$ during phase transformations. In line with experimental observations, our simulations successfully predict the geometric features of the finely twinned domains, and microstructural evolution pathways, and quantitatively estimate the macroscopic material response (e.g., voltage curves, stress distributions) during a galvanostatic discharge half cycle in Li$_2$Mn$_2$O$_4$. Additionally, our simulations reveal significant interfacial stresses at phase boundaries and particle surfaces, which suggest a potential mechanism for failure in Li$_{2x}$Mn$_2$O$_4$. Broadly, our results highlight the use of our modeling framework to investigate and design crystallographic microstructures in first-order phase transformation materials.

\section{Theory}\label{sec2}

\noindent In this section we use the Cauchy-Born Rule \citep{ericksen2008cauchy} and the Principle of Virtual Power \citep{gurtin2002gradient, anand2012cahn} and the Thermodynamic principles to derive the constitutive form of the diffusion-deformation theory. We start by analyzing the lattice deformations in Li$_2$Mn$_2$O$_4$, as a representative compound, and formulate a thermodynamically consistent framework to predict symmetry-lowering phase transformations in chemo-mechanical materials.

\subsection{Kinematics}

\noindent Consider a body occupying a region $\Omega$ in three-dimensional Euclidean space $\mathbb{R}^3$ in the reference configuration. Let $\mathbf{x}$ be the position of a point on $\Omega$ in this reference configuration. We describe the deformation of this body as a function $\boldsymbol{\chi}:\Omega \to \mathbb{R}^3$, in which $\boldsymbol{\chi}(\mathbf{x}, t)$ denotes the position of point $\mathbf{x}$ in the deformed configuration. Given any deformation $\boldsymbol{\chi}$, we describe the deformation gradient $\nabla\boldsymbol{\chi}$ as a matrix of partial derivatives with components,

\begin{eqnarray} \label{3.1} 
(\nabla\boldsymbol{\chi})_{iJ}=\frac{\partial \chi_i}{\partial x_J} \quad i,J = 1,2,3.
\end{eqnarray}

\noindent We use $\mathbf{F}$ to denote this deformation gradient, i.e., a rank-2 tensor $\mathbf{F}=\nabla\boldsymbol{\chi}$ and the gradient operator $\nabla$ is calculated with respect to the reference configuration. Please note, in this work, we denote vectors and tensors using bold lower-case and upper-case Roman letters, respectively. We denote the components of these vectors and tensors, with respect to a Cartesian basis, using upper-case (lower-case) indices in the reference (deformed) configuration. A vector with a hat is a unit normal vector (e.g., $\hat{\mathbf{n}}$) in the reference configuration.

\begin{figure}[ht!]
    \centering
    \includegraphics[width=0.8\textwidth]{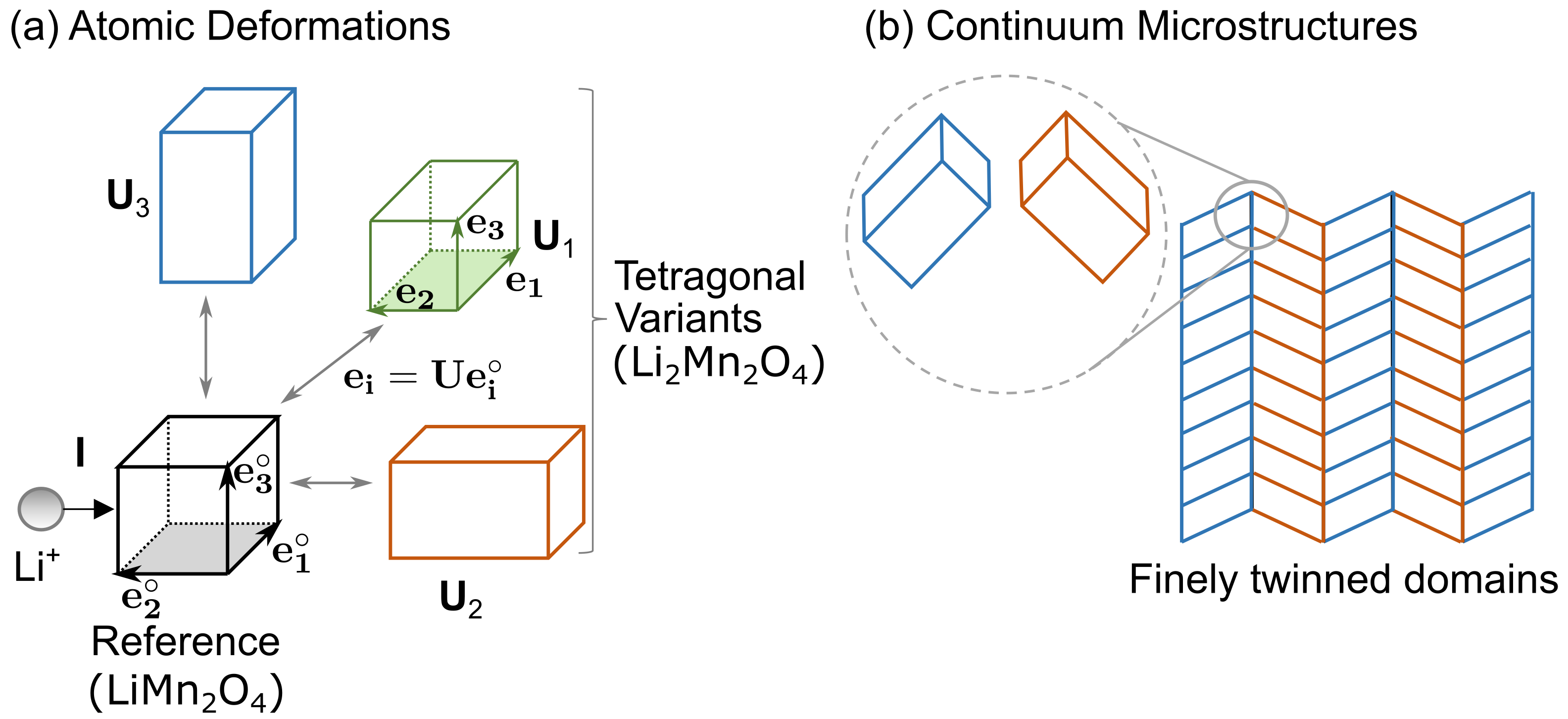}
    \caption{(a) A schematic illustration of the cubic-to-tetragonal lattice deformations in Li$_{(1-2)}$Mn$_2$O$_4$ at the atomic scale. The Cauchy-Born Rule states that the lattices deform according to the deformation gradient in Eq.~(\ref{cb2}). Here, the higher-symmetry cubic phase (LiMn$_2$O$_4$) is denoted by $\mathbf{I}$ and the lower-symmetry tetragonal phase (Li$_2$Mn$_2$O$_4$) are denoted by $\mathbf{U}_1,\mathbf{U}_2,\mathbf{U}_3$, respectively. (b) Individual tetragonal variants rotate and shear to form compatible twin interfaces. These interfaces are energy-minimizing deformations that form at the continuum level.}
    \label{Fig2}
\end{figure}

\subsection{Bravais Lattices}

\noindent In the continuum theory of crystalline solids, the Cauchy-Born rule, relates the movement of atoms in a crystal to the overall deformation of the body. That is, consider the body $\Omega$ as described above, and at each point $\mathbf{x} \in \Omega$, there is a Bravais lattice that defines the crystalline arrangement of atoms. This Bravais lattice is an infinite set of points in three-dimensional space that can be generated by the translation of a single point $\mathbf{o}$ through three linearly independent lattice vectors $\{\mathbf{e}_1, \mathbf{e}_2, \mathbf{e}_3\}$. In the reference configuration, the unit cell is defined by lattice vectors $\mathbf{e}^\circ_i$ and deforms according to the deformation gradient:
\begin{eqnarray} \label{cb2}
\mathbf{e}_i=\mathbf{F}\mathbf{e}^\circ_i.
\end{eqnarray}

\noindent We define the Green-Lagrange strain tensor as
\begin{eqnarray} \label{5.42}
\mathbf{E}
&=&\frac{1}{2}\left({\mathbf{F}}^\top \mathbf{F}-\mathbf{I}\right).
\end{eqnarray}

\noindent in which $\mathbf{I}$ is the identity matrix. The strain tensor in coordinate notation is given by $E_{I J} =  \frac{1}{2}(F_{kI}F_{kJ}-\delta_{{I J}})$.

\vspace{2mm}
\noindent The Cauchy-Born rule gives an exact correspondence between these structural transformations of individual lattices and continuum microstructures at a material point. For example, in first-order phase transformation materials (i.e., materials undergoing displacive-type of transformation), we describe the structural transformation of lattices using a positive-definite symmetric matrix $\mathbf{U}$ 
\begin{eqnarray}
\mathbf{e}_i=\mathbf{U}\mathbf{e}^\circ_i.
\end{eqnarray}

\noindent Fig.~\ref{Fig2}(a) shows a cubic-to-tetragonal deformation of lattices with the three tetragonal variants described by $\mathbf{U}_1$, $\mathbf{U}_2$ and $\mathbf{U}_3$, respectively. In intercalation materials, such as Li$_2$Mn$_2$O$_4$, a similar displacive-type of lattice transformation is observed using in-situ TEM.

\subsection{Compatibility Conditions}\label{sec2.3}

\begin{figure}[ht!]
    \centering
    \includegraphics[width=0.8\textwidth]{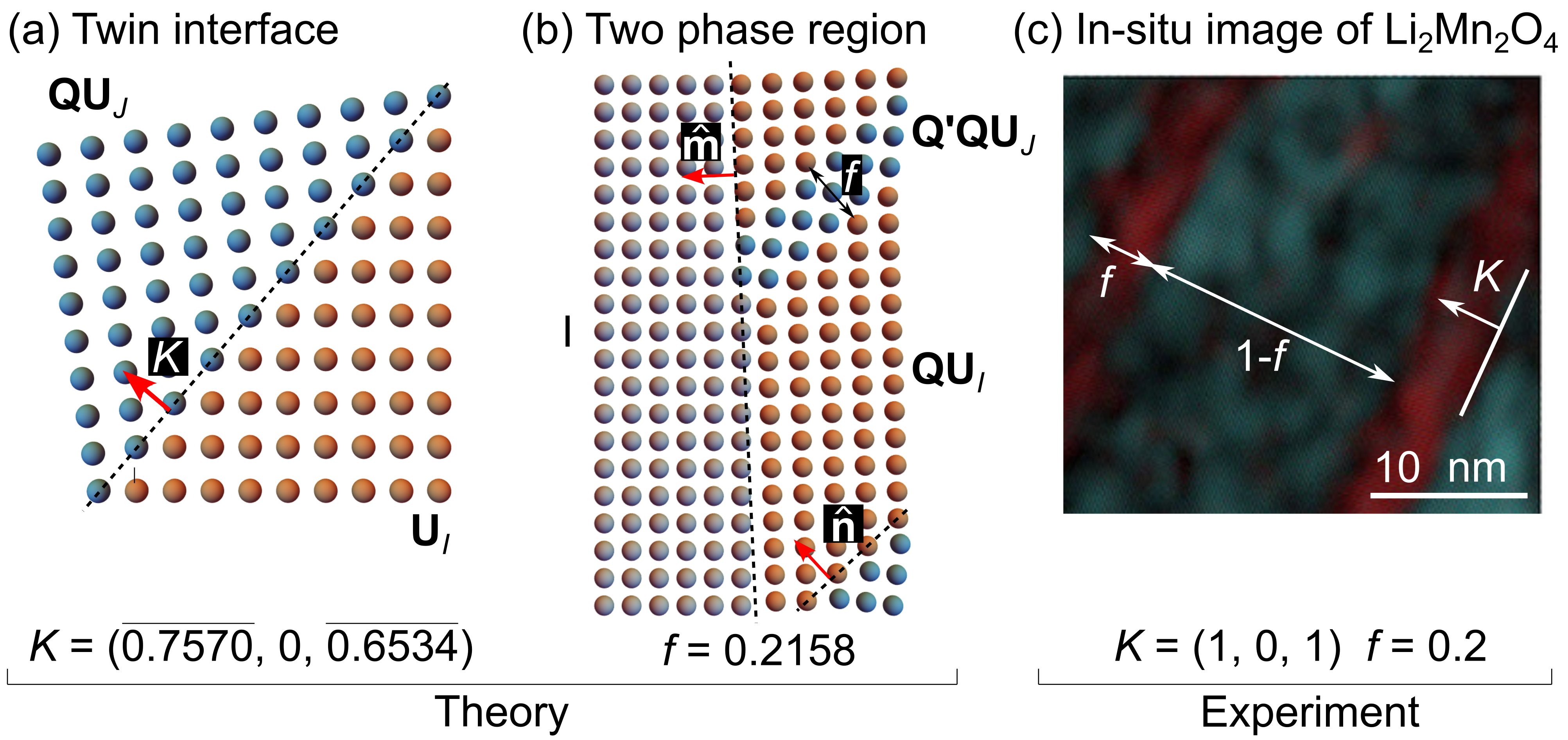}
    \caption{We compare the analytical construction of twin interfaces with HRTEM images of Li$_{2x}$Mn$_{2}$O$_4$ \citep{erichsen2020tracking}. Using Eq.~(\ref{eq: twin interface}) we analytically construct the twin interface between two Li$_2$Mn$_2$O$_4$ tetragonal variants. A cross-sectional view of the twin interface shows a twin-plane $\mathbf{K} = (0.7570, 0, 0.6534)$ \citep{zhang2023designing} that compares favorably with the experimental measurement from \citep{erichsen2020tracking}. (b) Using Eq.~(\ref{eq: AM interface}) we analytically construct the LiMn$_2$O$_4$/Li$_2$Mn$_2$O$_4$ interface. Our calculations predict a volume fraction of the twinned mixture to be $f = 0.2158$ \citep{zhang2023designing} that compares favorably with the experimental measurement from \citep{erichsen2020tracking}. (c) Bragg-filtered HRTEM image of the lamellar microstructures showing (101) twining plane in Cartesian coordinates and volume fraction $f = 0.2$ \citep{erichsen2020tracking} (Reproduced with permission from American Chemical Society).}
    \label{Fig3}
\end{figure}

\noindent During phase transformation, the tetragonal variants rotate and fit compatibly with each other forming energy-minimizing deformations called twin interfaces. That is, two lattice variants described by transformation matrices $\mathbf{U}_I$ and $\mathbf{U}_J$ satisfy the Hadamard jump condition (or the Kinematic compatibility condition):
\begin{equation}
    \mathbf{QU}_{J} - \mathbf{U}_{I} = \mathbf{a}\otimes\mathbf{\hat{n}}.
    \label{eq: twin interface}
\end{equation}

\noindent for a given rotation matrix $\mathbf{Q}$ and vectors $\mathbf{a}\neq0$ and $\mathbf{\hat{n}}$, see Fig. \ref{Fig3}. The solution to Eq. (\ref{eq: twin interface}) then describes a twin plane that connects the two lattice variants $\mathbf{U}_I$ and $\mathbf{U}_J$ coherently. These twin interfaces have been observed in intercalation materials, such as Li$_2$Mn$_2$O$_4$, in-situ, during battery operation. These twin interfaces are rarely found in isolation. As illustrated in Fig. \ref{Fig3}, finely twinned microstructures form when any two lattice variants, with deformation tensors $\mathbf{U}_I$ and $\mathbf{U}_J$, satisfy the kinematic compatibility condition for some scalar $0 \leq f \leq 1$\citep{ball1987fine, bhattacharya2003microstructure}:

\begin{equation}
    \mathbf{Q}'(f\mathbf{Q}\mathbf{U}_J+(1-f)\mathbf{U}_I)=\mathbf{I}+\mathbf{b}\otimes\mathbf{\hat{m}}.
    \label{eq: AM interface}
\end{equation}

\noindent Here the reference cubic lattice is represented by $\mathbf{I}$, rotation matrices by $\mathbf{Q}$ and $\mathbf{Q}'$, and vectors $\mathbf{b} \neq 0$ and $\mathbf{\hat{m}}$. These martensite-like microstructures describe energy-minimizing deformations that reduce coherency stresses at the phase boundary and, arise from the multi-well structure of the energy landscape. We discuss this in detail in section \ref{sec:Microstructure Evolution}. 

\vspace{2mm}
\noindent In our recent work \citep{zhang2023designing}, we used compatibility conditions in Eqs.~(\ref{eq: twin interface}) and (\ref{eq: AM interface}) to analytically construct the twin interfaces in Li$_{2x}$Mn$_2$O$_4$. As shown in Fig.~\ref{Fig3}, our analytical solutions of both the twin plane $\mathbf{K}$ and the volume fraction $f$ for the austenite-martensite microstructure are consistent with experimental measurements \citep{erichsen2020tracking}. These observations further support our efforts to derive a diffusion-deformation model to predict crystallographic microstructures in Li$_2$Mn$_2$O$_4$.

\subsection{Mass Balance} 

\noindent Consider any spatial region $\mathcal{P}$ inside the reference body $\Omega$, with an outward normal $\hat{\mathbf{n}}$ and a boundary $\partial \mathcal{P}$. The diffusion of species (e.g., Li-ions in batteries) across this boundary $\partial \mathcal{P}$ is accompanied by a change in the species concentration $c(\mathbf{x},t)$ and characterized by a flux $\mathbf j(\mathbf{x},t)$. That is, the rate of change of the chemical species across $\mathcal{P}$ is given by
\begin{eqnarray} 
\int_{\mathcal{P}} \dot c  ~dV = - \int_{\partial \mathcal{P}} \mathbf j  \cdot \hat{\mathbf n} ~dA.
\end{eqnarray}

\noindent Using the Divergence theorem over the integral we have,
\begin{eqnarray} 
\int_{\partial \mathcal{P}} (\dot c  + \nabla \cdot \mathbf j) ~dV =0.
\end{eqnarray}

\noindent Note that the choice of $\mathcal{P}$ was arbitrary, and therefore the local mass balance law is as follows
\begin{eqnarray} \label{13}
\dot c  = -\nabla \cdot \mathbf j.
\end{eqnarray}

\subsection{Force Balance} 
% derived from the principle of virtual power
\noindent The principle of virtual power states that there exists a fundamental power balance between the external power $W_{\mathrm{ext}}(\mathcal{P})$ expended on $\mathcal{P}$, and the internal power $W_{\mathrm{int}}(\mathcal{P})$ expended within $\mathcal{P}$. To obtain the expressions for external and internal powers, we follow \cite{gurtin2002gradient} and \cite{anand2012cahn} and use ``rate-like" kinematic descriptors --- $\dot {\boldsymbol{\chi}}$, $\dot {\mathbf{F}}$, $\nabla \dot {\mathbf{F}}$, $\dot c$,  and $\nabla \dot c$ --- to derive the associated balance of forces by the principle of virtual power. These rates are not independent but are constrained by:
\begin{eqnarray} \label{14}
\dot {\mathbf{F}} =\nabla  \dot {\boldsymbol{\chi}}.
\end{eqnarray}

\begin{longtable}{p{0.3\textwidth}p{0.13\textwidth}p{0.3\textwidth}p{0.13\textwidth}}
  \caption{Individual force systems and their power conjugates} \\ 
  \label{TPVP}\\
   \multicolumn{4}{l}{External Power  \hspace{0.31\textwidth} Internal Power} \\
  \hline
	  Force &  Power & Force & Power\\
  & Conjugate & & Conjugate\\
    \hline
    Traction $\mathbf t$ &  $\dot {\boldsymbol{\chi}}$ & Stress $\mathbf{T}_{\mathrm{R}}$ & $\dot {\mathbf{F}}$ \\
    Moment $\mathbf m$ & $(\nabla\dot {\boldsymbol{\chi}})\hat{\mathbf{n}}$ & Higher-order stress $\mathbf{Y}$ & $ \nabla \dot {\mathbf{F}}$ \\
    Line force $\mathbf l$ & $\dot {\boldsymbol{\chi}}$ & Scalar microscopic force $\pi$ & $\dot c$\\
    Body force $\mathbf{b}$ & $\dot {\boldsymbol{\chi}}$ & Vector microscopic force $\boldsymbol{\xi}$ & $\nabla \dot c$\\
    Scalar microscopic traction $\zeta$ & $\dot c$ & & \\    
    \hline
    \label{Tab:1}
\end{longtable}

\noindent In Table~\ref{Tab:1} we identify individual force systems expending power externally on $\mathcal{P}$ and expending power internally within $\mathcal{P}$. We use these individual force systems to construct the total external $W_{\mathrm{ext}}(\mathcal{P})$ and internal power $W_{\mathrm{int}}(\mathcal{P})$ as follows:
\begin{align}\label{15}
W_{\mathrm{ext}}(\mathcal{P}) &=\int_{\partial \mathcal{P}} \mathbf t  \cdot \dot {\boldsymbol{\chi}}  ~dA + \int_{\partial \mathcal{P}} \mathbf m  \cdot (\nabla\dot {\boldsymbol{\chi}})\hat{\mathbf{n}}  ~dA + \int_{\zeta^L} \mathbf l  \cdot \dot {\boldsymbol{\chi}}  ~dL + \int_{\mathcal{P}}  \mathbf{b} \cdot \dot {\boldsymbol{\chi}}~dV + \int_{\partial \mathcal{P}} \zeta  \dot c ~dA,\nonumber\\
W_{\mathrm{int}}(\mathcal{P}) &=\int_{\mathcal{P}} (\mathbf{T}_{\mathrm{R}} \colon \dot {\mathbf{F}} + \mathbf{Y}~\vdots~\nabla \dot {\mathbf{F}}+ \pi \dot c + \boldsymbol{\xi} \cdot \nabla \dot c)  ~dV.
\end{align} 

\noindent Please note that in Eq.~(\ref{15}) the stresses $\mathbf{T}_{\mathrm{R}}$, $\mathbf{Y}$, and microscopic forces $\pi$ and $\boldsymbol{\xi}$ are defined over the body at all times. We define a line force $\mathbf{l}$ across the boundary edge $\zeta^L$ of the region $\mathcal{P}$. We assume that at any given time, the fields $\boldsymbol{\chi}$, $\mathbf{F}$, and $c$ are known, and we consider the fields $\dot {\boldsymbol{\chi}}$, $\dot {\mathbf{F}}$, and $\dot {c}$ as virtual velocities. We specify each of these virtual velocities independently and keep them consistent with Eq.~(\ref{14}). Next, we introduce virtual fields of the form $\widetilde{\boldsymbol{\chi}}$, $\widetilde{\mathbf{F}}$, and $\widetilde{c}$, to distinguish these fields from the previously described virtual velocities (as associated with the real-time evolution of the body) and we ensure that virtual fields satisfy:
\begin{eqnarray} \label{17}
\widetilde{\mathbf{F}} =\nabla  \widetilde{\boldsymbol{\chi}}.
\end{eqnarray}

\noindent We define a generalized virtual velocity list $\mathcal V =(\widetilde{\boldsymbol{\chi}}, \widetilde{\mathbf{F}}, \widetilde{c})$ that is consistent with Eq. (\ref{17}). Following \cite{anand2012cahn} and  \cite{gurtin2002gradient}, we assume that under a change in the frame, the fields comprising the generalized virtual velocity convert as their nonvirtual counterparts. That is,
\begin{eqnarray} \label{18}
\widetilde{\mathbf{F}}^{\ast} =\mathbf{Q}\widetilde{\mathbf{F}} + \dot {\mathbf{Q}}\mathbf{F}.
\end{eqnarray}

\noindent The external and internal power expenditures, respectively, are:
\begin{eqnarray} \label{19}
W_{\mathrm{ext}}(\mathcal{P}) &=&\int_{\partial \mathcal{P}} \mathbf t  \cdot \widetilde{\boldsymbol{\chi}}  ~dA + \int_{\partial \mathcal{P}} \mathbf m  \cdot (\nabla\widetilde {\boldsymbol{\chi}})\hat{\mathbf{n}}  ~dA + \int_{\zeta^L} \mathbf l  \cdot \widetilde {\boldsymbol{\chi}}  ~dL + \int_{\mathcal{P}}  \mathbf{b} \cdot \widetilde {\boldsymbol{\chi}}  ~dV+ \int_{\partial \mathcal{P}} \zeta  \widetilde c ~dA,
\end{eqnarray}
\begin{eqnarray} \label{20}
W_{\mathrm{int}}(\mathcal{P}) &=&\int_{\mathcal{P}} (\mathbf{T}_{\mathrm{R}} \colon \widetilde {\mathbf{F}} + \mathbf{Y}~\vdots~\nabla \widetilde {\mathbf{F}}+ \pi \widetilde c + \boldsymbol{\xi} \cdot \nabla \widetilde c)  ~dV.
\end{eqnarray}

\noindent From Eqs.~(\ref{19}) and (\ref{20}), the principle of virtual power comprises two essential requirements:\\

\noindent \textit{Power Balance}: Given any spatial region $\mathcal{P}$,
\begin{eqnarray} \label{21}
W_{\mathrm{ext}}(\mathcal{P}, \mathcal V) =W_{\mathrm{int}}(\mathcal{P}, \mathcal V)  \ \text{for all generalized virtual velocities $\mathcal V$}.
\end{eqnarray}

\noindent \textit{Frame-Indifference}: Given any spatial region $\mathcal{P}$ and any generalized virtual velocity  $\mathcal V$,
\begin{eqnarray} \label{22}
W_{\mathrm{int}}(\mathcal{P}, \mathcal V)  \ \text{is invariant for all changes in frame}.
\end{eqnarray}

\noindent To deduce the outcomes of the principle of virtual power, we assume that  Eqs.~(\ref{21}) and (\ref{22}) are satisfied. We note that in applying the virtual balance described by Eq.~(\ref{21}), we are free to choose any $\mathcal V$ that is consistent with the constraint Eq.~(\ref{17}). 

\subsubsection{Macroscopic Force and Moment Balances}\label{sec:2.5.1}

\noindent Let us assume $\widetilde{c}=0$. For this choice of the generalized virtual velocity $\mathcal V$, the constraint in Eq.~(\ref{17}) and the Power Balance requirement in Eq.~(\ref{21}) yield:
\begin{eqnarray} \label{23}
\int_{\partial \mathcal{P}} \mathbf t  \cdot \widetilde{\boldsymbol{\chi}}  ~dA + \int_{\partial \mathcal{P}} \mathbf m  \cdot (\nabla\widetilde {\boldsymbol{\chi}})\hat{\mathbf{n}}  ~dA + \int_{\zeta^L} \mathbf l  \cdot \widetilde{\boldsymbol{\chi}}  ~dL + \int_{\mathcal{P}}  \mathbf{b} \cdot \widetilde {\boldsymbol{\chi}}  ~dV =
 \int_{\mathcal{P}} (\mathbf{T}_{\mathrm{R}} \colon \nabla  \widetilde{\boldsymbol{\chi}} + \mathbf{Y}~\vdots~\nabla  \nabla  \widetilde{\boldsymbol{\chi}})  ~dV.
\end{eqnarray}

\noindent We systematically simplify Eq.~(\ref{23}) (see the detailed steps in \ref{A}) and derive the macroscopic force balance. For our purposes, we list the form of this force balance
\begin{eqnarray}\label{24}
0&=&-\int _{\mathcal{P}} \widetilde{\boldsymbol{\chi}} \cdot (\nabla \cdot \mathbf{T}_{\mathrm{R}}^{\top}-\nabla \cdot(\nabla \cdot \mathbf{Y}^{\top})^{\top}+\mathbf{b}) ~dV \nonumber \\ 
&&+ \int_{\partial \mathcal{P}} \widetilde{\boldsymbol{\chi}} \cdot \left(\mathbf{T}_{\mathrm{R}}\hat{\mathbf{n}}-(\nabla \cdot \mathbf{Y}^{\top})\hat{\mathbf{n}}-\nabla^s \cdot (\mathbf{Y} \cdot \hat{\mathbf{n}})^{\top}- \mathbf{Y}\colon ((\nabla^s \cdot \hat{\mathbf{n}})\hat{\mathbf{n}} \otimes \hat{\mathbf{n}}-\nabla^s \hat{\mathbf{n}})-\mathbf{t}\right)  ~dA \nonumber \\ 
&& +  \int_{\partial \mathcal{P}} (\nabla\widetilde {\boldsymbol{\chi}})\hat{\mathbf{n}} \cdot (\mathbf{Y}\colon(\hat{\mathbf{n}} \otimes \hat{\mathbf{n}})-\mathbf{m})~dA 
+ \int_{\zeta^L} \widetilde{\chi}_i (\left[\left[\hat{n}^{\mathit{\Gamma}}_J \hat{n}_K Y_{iJK}  \right] \right]-l_i)~dL.
\end{eqnarray}

\noindent Here $\nabla^s =(\mathbf{I}-\hat{\mathbf{n}}\otimes\hat{\mathbf{n}})\nabla$ is a surface gradient operator and $\mathit{\Gamma} = \partial\mathcal{P}$. Eq.~(\ref{24}) holds for all choices of $\mathcal{P}$ and $\widetilde{\boldsymbol{\chi}}$ and, the standard variational arguments, respectively, lead to the following traction conditions:
\begin{eqnarray}\label{25}
\mathbf{T}_{\mathrm{R}}\hat{\mathbf{n}}-(\nabla \cdot \mathbf{Y}^{\top})\hat{\mathbf{n}}-\nabla^s \cdot (\mathbf{Y} \cdot \hat{\mathbf{n}})^{\top}-\mathbf{Y}\colon ((\nabla^s \cdot \hat{\mathbf{n}})\hat{\mathbf{n}} \otimes \hat{\mathbf{n}}-\nabla^s \hat{\mathbf{n}})&=&\mathbf{t},\nonumber\\
\mathbf{Y}\colon(\hat{\mathbf{n}} \otimes \hat{\mathbf{n}})&=&\mathbf{m},\nonumber\\
\left[\left[\hat{n}^{\mathit{\Gamma}}_J \hat{n}_K Y_{iJK}  \right] \right]&=&l_i~~\text{with}~~\mathit{\Gamma} = \partial \mathcal{P}.
\end{eqnarray}

\noindent and the local macroscopic force balance:
\begin{eqnarray} \label{28}
  \nabla \cdot \mathbf{T}_{\mathrm{R}}^{\top}-\nabla \cdot(\nabla \cdot \mathbf{Y}^{\top})^{\top}+\mathbf{b}=0.
\end{eqnarray}

\noindent The principle of frame-indifference Eq.~(\ref{22}) (underlying most physical laws including the principle of virtual power) requires that the internal power is invariant to all changes of frame :
\begin{eqnarray} \label{29}
W^{\ast}_{\mathrm{int}}(\mathcal{P}, \mathcal V^{\ast}) = W_{\mathrm{int}}(\mathcal{P}, \mathcal V),
\end{eqnarray}

\noindent in which $\mathcal V^{\ast}$ is the generalized virtual velocity in the new frame. In this new frame, $\boldsymbol{\xi}$ is transformed to $\boldsymbol{\xi}^{\ast}$, and $\widetilde{\mathbf{F}}, \nabla \widetilde{\mathbf{F}}$ are transformed according to Eq.~(\ref{18}). Scalar fields, such as $\widetilde{c}$ and $\pi$, are frame invariant and, gradient fields, such as $\nabla \widetilde{c}$ is equal to  $(\nabla \widetilde{c})^{\ast}$. With these transformations, a change in frame converts internal power $W_{int}(\mathcal{P}, \mathcal V)$ to
\begin{eqnarray} \label{30}
W^{\ast}_{\mathrm{int}}(\mathcal{P}, \mathcal V^{\ast}) &=&\int_{\mathcal{P}} (\mathbf{T}_{\mathrm{R}}^{\ast} \colon (\mathbf{Q} \widetilde{\mathbf{F}} + \dot {\mathbf{Q}}\mathbf{F}) + \mathbf{Y}^{\ast}~\vdots~(\nabla (\mathbf{Q} \widetilde{\mathbf{F}}) +  \nabla (\dot {\mathbf{Q}} \mathbf{F}))+ \pi \widetilde c + \boldsymbol{\xi}^{\ast} \cdot \nabla \widetilde c)  ~dV \nonumber\\
&=&\int_{\mathcal{P}} (\mathbf{Q}^{\top} \mathbf{T}_{\mathrm{R}}^{\ast} \colon ( \widetilde{\mathbf{F}} +\mathbf{Q}^{\top} \dot {\mathbf{Q}}\mathbf{F}) + \mathbf{Y}^{\ast}~\vdots~(\nabla (\mathbf{Q} \widetilde{\mathbf{F}}) +  \nabla (\dot {\mathbf{Q}} \mathbf{F})) + \pi \widetilde c + \boldsymbol{\xi}^{\ast} \cdot \nabla \widetilde c)  ~dV.
\end{eqnarray}

\noindent The frame-indifference requirement in Eqs.~(\ref{29}) and (\ref{30}) yields
\begin{eqnarray} 
\int_{\mathcal{P}} (\mathbf{Q}^{\top} \mathbf{T}_{\mathrm{R}}^{\ast} \colon (\widetilde{\mathbf{F}} +\mathbf{Q}^{\top} \dot {\mathbf{Q}}\mathbf{F}) +  \mathbf{Y}^{\ast}~\vdots~(\nabla (\mathbf{Q} \widetilde{\mathbf{F}}) +  \nabla (\dot {\mathbf{Q}} \mathbf{F}))
+ \pi \widetilde c + \boldsymbol{\xi}^{\ast} \cdot \nabla \widetilde c)  ~dV \nonumber\\
 =
\int_{\mathcal{P}} (\mathbf{T}_{\mathrm{R}} \colon \widetilde {\mathbf{F}} + \mathbf{Y}~\vdots~\nabla \widetilde {\mathbf{F}}+ \pi \widetilde c + \boldsymbol{\xi} \cdot \nabla \widetilde c)  ~dV.
\end{eqnarray}

\noindent Since $\mathcal{P}$ is an arbitrary part of the reference body $\Omega$, we have
\begin{eqnarray}
\mathbf{Q}^{\top} \mathbf{T}_{\mathrm{R}}^{\ast} \colon ( \widetilde{\mathbf{F}} +\mathbf{Q}^{\top} \dot {\mathbf{Q}}\mathbf{F}) +  \mathbf{Y}^{\ast}~\vdots~(\nabla (\mathbf{Q} \widetilde{\mathbf{F}}) +  \nabla (\dot {\mathbf{Q}} \mathbf{F}))  + \boldsymbol{\xi}^{\ast} \cdot \nabla \widetilde c
 =
\mathbf{T}_{\mathrm{R}} \colon \widetilde {\mathbf{F}} + \mathbf{Y}~\vdots~\nabla \widetilde {\mathbf{F}} + \boldsymbol{\xi} \cdot \nabla \widetilde c.
\end{eqnarray}

\noindent Note that this change in frame can be arbitrary and we can proceed in two ways. First, we assume a time-independent rotation, $\dot {\mathbf{Q}}=0$, and we have:
\begin{eqnarray} \label{31}
(\mathbf{T}_{\mathrm{R}} - (\mathbf{Q}^{\top} \mathbf{T}_{\mathrm{R}}^{\ast})) \colon \widetilde {\mathbf{F}} +\mathbf{Y}~\vdots~\nabla \widetilde {\mathbf{F}} -  \mathbf{Y}^{\ast}~\vdots~\nabla (\mathbf{Q} \widetilde{\mathbf{F}})  + (\boldsymbol{\xi} - \boldsymbol{\xi}^{\ast}) \cdot \nabla \widetilde c = 0.
\end{eqnarray}

\noindent As Eq.~(\ref{31}) must hold for all $\widetilde {\mathbf{F}}$, $\nabla \widetilde {\mathbf{F}}$, and $\nabla \widetilde c$, the stresses $\mathbf{T}_{\mathrm{R}}$ and $\mathbf{Y}$ need to satisfy
\begin{eqnarray}
\mathbf{T}_{\mathrm{R}}^{\ast} = \mathbf{Q} \mathbf{T}_{\mathrm{R}} \ \  \mathrm{and} \ \  \mathbf{Y}~\vdots~\nabla \widetilde {\mathbf{F}} =  \mathbf{Y}^{\ast}~\vdots~\nabla (\mathbf{Q} \widetilde{\mathbf{F}}).
\end{eqnarray}

\noindent and the microstress $\boldsymbol{\xi}$ is invariant
\begin{eqnarray}
\boldsymbol{\xi}^{\ast} = \boldsymbol{\xi}.
\end{eqnarray}

\noindent Second, we assume a change of frame satisfying $\mathbf{Q}=\mathbf{I}$, such that $\dot {\mathbf{Q}}$ is an arbitrary skew tensor. With this assumption, we obtain
\begin{eqnarray} \label{37}
\mathbf{T}_{\mathrm{R}} \mathbf{F}^{\top} \colon  \dot {\mathbf{Q}} + \mathbf{Y}~\vdots~\nabla(\dot {\mathbf{Q}}\mathbf{F}) =0
\end{eqnarray}

\noindent and the stress $\mathbf{T}_{\mathrm{R}} \mathbf{F}^{\top}$ is symmetric,
\begin{eqnarray} \label{38}
\mathbf{T}_{\mathrm{R}} \mathbf{F}^{\top} &=& \mathbf{F} \mathbf{T}_{\mathrm{R}}^{\top}.
\end{eqnarray}

\noindent Here $\mathbf{T}_{\mathrm{R}}$ represents the classical first Piola-Kirchhoff stress tensor, $\mathbf{Y}$ represents the third-order
stress tensor that is conjugate to the higher-order deformation gradient. Eq.~(\ref{28}) and Eq.~(\ref{38}) represent the local macroscopic force and moment balances in the reference body.

\subsubsection{Microscopic Force Balance} 
\noindent We next analyze the microscopic counterparts of the macroscopic force balance. We consider a generalized virtual velocity with $\widetilde{\boldsymbol{\chi}} = \mathbf{0}$ and an arbitrary virtual field $\widetilde{c}$. By substituting these fields into the external and internal power expenditures, Eqs.~(\ref{19}) and (\ref{20}), and together with the power balance requirement of the principle of virtual power Eq.~(\ref{21}) and the constraint Eq.~(\ref{17}), we have the following microscopic virtual-power relation:
\begin{eqnarray} \label{39}
 \int_{\partial \mathcal{P}} \zeta  \widetilde c ~dA =\int_{\mathcal{P}} (\pi \widetilde c + \boldsymbol{\xi} \cdot \nabla \widetilde c)  ~dV.
\end{eqnarray}

\noindent Eq.~(\ref{39}) holds for all values of $\widetilde c$ and for all choices of $\mathcal{P}$. By applying the divergence theorem, we have
\begin{eqnarray} \label{40}
 \int_{\partial \mathcal{P}} (\zeta -  \boldsymbol{\xi} \cdot \hat{\mathbf n})  \widetilde c ~dA =\int_{\mathcal{P}} (\pi  - \nabla \cdot \boldsymbol{\xi} )\widetilde c ~dV.
\end{eqnarray}

\noindent Eq.~(\ref{40}) must hold for all choices of $\mathcal{P}$ and all $\widetilde c$, standard variational arguments yield the microscopic traction condition:
\begin{eqnarray} \label{41}
\zeta = \boldsymbol{\xi} \cdot \hat{\mathbf n},
\end{eqnarray}

\noindent and the microscopic force balance:
\begin{eqnarray} \label{42}
\pi  = \nabla \cdot \boldsymbol{\xi}.
\end{eqnarray}

\noindent Overall, Eqs.~(\ref{25}), (\ref{28}), (\ref{38}), (\ref{41}), and (\ref{42}) represent the principle of virtual power.

\subsection{Imbalance of Energy} \label{Free energy imbalance}
\noindent We next derive the free energy imbalance under isothermal conditions. Let $\psi$ represent the Helmholtz free energy density of the system per unit volume and $\mu$ represent the chemical potential of the diffusing species in the reference configuration. By neglecting the kinetic energy of the system, we have:

\begin{eqnarray} \label{43}
 \frac{d}{dt}\left(\int_{\mathcal{P}} \psi~dV\right) \leq W_{\mathrm{ext}}(\mathcal{P}) -  \int_{\partial \mathcal{P}} \mu \mathbf j  \cdot \hat{\mathbf n}~dA.
\end{eqnarray}

\noindent Using the power balance property $W_{\mathrm{ext}}(\mathcal{P}) =W_{\mathrm{int}}(\mathcal{P})$, Eq.~(\ref{15}) and the divergence theorem, we have:
\begin{eqnarray} \label{44}
 \int_{\mathcal{P}} (\dot {\psi} - \mathbf{T}_{\mathrm{R}} \colon \dot {\mathbf{F}} - \mathbf{Y}~\vdots~\nabla \dot {\mathbf{F}}- \pi \dot c - \boldsymbol{\xi} \cdot \nabla \dot c + \mu  \nabla \cdot  \mathbf j  +   \mathbf j  \cdot \nabla  \mu)  ~dV  \leq 0.
\end{eqnarray}

\noindent Using the mass balance from Eq.~(\ref{13}) and noting that Eq.~(\ref{44}) holds for all spatial regions $\mathcal{P}$, we write the local form of the free energy imbalance as:
\begin{eqnarray} \label{45}
\dot {\psi} - \mathbf{T}_{\mathrm{R}} \colon \dot {\mathbf{F}} - \mathbf{Y}~\vdots~\nabla \dot {\mathbf{F}}- \mu_{net} \dot c - \boldsymbol{\xi} \cdot \nabla \dot c  +   \mathbf j  \cdot \nabla  \mu  \leq 0.
\end{eqnarray}

\noindent In Eq.~(\ref{45}), we introduce a net chemical potential, 
\begin{eqnarray} \label{46}
\mu_{net} = \mu + \pi.
\end{eqnarray}

\noindent At this point, we define a dissipation density $\mathcal D$ per unit volume per unit time:
\begin{eqnarray} \label{47}
\mathcal D = \mathbf{T}_{\mathrm{R}} \colon \dot {\mathbf{F}} + \mathbf{Y}~\vdots~\nabla \dot {\mathbf{F}} + \mu_{net} \dot c + \boldsymbol{\xi} \cdot \nabla \dot c  -   \mathbf j  \cdot \nabla  \mu - \dot {\psi} \geq 0.
\end{eqnarray}

\noindent Please note that all quantities in Eq.~(\ref{45}) and Eq.~(\ref{47}) are invariant under a change in frame.

\subsection{Constitutive Theory} 
\noindent We next consider the constitutive forms for the free energy density $\psi$, the first Piola-Kirchhoff stress tensor $\mathbf{T}_{\mathrm{R}}$, the third-order stress tensor $\mathbf{Y}$, the net chemical potential $\mu_{net}$, and the vector microscopic force $\boldsymbol{\xi}$, and the species flux $\mathbf j$. Following the free energy imbalance derived in Eq.~(\ref{45}), we note that each term can be expressed as a functional of the deformation gradient $\mathbf{F}, \nabla\mathbf{F}$, species concentration $c,\nabla c$, and the chemical potential $\nabla \mu$:
\begin{eqnarray} \label{48}
\psi &=& \hat{\psi} (\mathbf{F}, \nabla \mathbf{F},  c,  \nabla c),\nonumber\\
\mathbf{T}_{\mathrm{R}} &=& \hat{\mathbf{T}}_{\mathrm{R}} (\mathbf{F}, \nabla \mathbf{F},  c,  \nabla c),\nonumber\\
\mathbf{Y} &=&\hat{\mathbf{Y}} (\mathbf{F}, \nabla \mathbf{F},  c,  \nabla c),\nonumber\\
\mu_{net} &=& \hat{\mu}_{net} (\mathbf{F}, \nabla \mathbf{F},  c,  \nabla c),\nonumber\\
\boldsymbol{\xi} &=&\hat{\boldsymbol{\xi}} (\mathbf{F}, \nabla \mathbf{F},  c,  \nabla c),\nonumber\\
\mathbf j&=& \hat{\mathbf j} (\mathbf{F}, \nabla \mathbf{F},  c,  \nabla c, \nabla  \mu).
\end{eqnarray}

\noindent Substituting the constitutive forms in Eq.~(\ref{48}) into the free-energy imbalance in Eq.~(\ref{45}), and using the chain rule, we have
\begin{eqnarray} \label{50}
[\frac{\partial {\hat\psi}}{\partial \mathbf{F}}-\mathbf{T}_{\mathrm{R}}] \colon \dot {\mathbf{F}}
+[\frac{\partial \hat{\psi}}{\partial \nabla \mathbf{F}}- \mathbf{Y}]~\vdots~\nabla \dot {\mathbf{F}}
+[\frac{\partial \hat{\psi}}{\partial  c}-\mu_{net}] \dot {c}+ 
[\frac{\partial \hat{\psi}}{\partial \nabla c} - \boldsymbol{\xi}]  \cdot \nabla \dot c
+ \hat{\mathbf j}  \cdot \nabla  \mu \leq 0.
\end{eqnarray}

\noindent This above inequality holds for all values of $\mathbf{F}$, $\nabla \mathbf{F}$, $c$, and $\nabla c$, which, in Eq.~(\ref{50}) appear in a linear form. The corresponding coefficients of these fields must vanish, so that $\dot {\mathbf{F}}$, $ \nabla \dot {\mathbf{F}}$, $\dot {c}$, and $\nabla \dot c$ can not be chosen to violate the free energy imbalance in Eq.~(\ref{50}). This argument, consequently, leads to the following thermodynamic restriction in which the first Piola-Kirchhoff stress $\mathbf{T}_{\mathrm{R}}$, the third-order stress $\mathbf{Y}$, the chemical potential $\mu$, and the vector microscopic force $\boldsymbol{\xi}$ are defined as derivatives of the free energy function $\hat\psi$:
\begin{eqnarray} \label{51}
\mathbf{T}_{\mathrm{R}} &=&\frac{\partial \hat{\psi} (\mathbf{F}, \nabla \mathbf{F},  c,  \nabla c)}{\partial \mathbf{F}},
\nonumber\\
\mathbf{Y} &=&\frac{\partial \hat{\psi} (\mathbf{F}, \nabla \mathbf{F},  c,  \nabla c)}{\partial \nabla \mathbf{F}},
\nonumber\\
\mu &=&\frac{\partial \hat{\psi} (\mathbf{F}, \nabla \mathbf{F},  c,  \nabla c)}{\partial  c} - \pi,\nonumber\\
\boldsymbol{\xi} &=&\frac{\partial \hat{\psi} (\mathbf{F}, \nabla \mathbf{F},  c,  \nabla c)}{\partial \nabla c}.
\end{eqnarray}

\noindent In Eq.~(\ref{51}), please note that the symmetry condition for third-order stress $\mathbf{Y}$ is $Y_{iJK} = Y_{iKJ}$. The dissipation inequality introduced in Eq.~(\ref{47}) reduces to
\begin{eqnarray} \label{52}
\mathcal D = - \hat{\mathbf j} (\mathbf{F}, \nabla \mathbf{F},  c,  \nabla c, \nabla  \mu)  \cdot \nabla  \mu  \geq 0.
\end{eqnarray}

\noindent Recalling the definition of $\mu_{net} = \mu + \pi$ in Eq.~(\ref{46}) and by using the constitutive form for $\mu$ in Eq.~(\ref{51}) and the microforce balance for $\pi$ in Eq.~(\ref{42}), we derive the constitutive equation for the chemical potential:
\begin{eqnarray} \label{53}
\mu =\frac{\partial \hat{\psi}(\mathbf{F}, \nabla \mathbf{F},  c,  \nabla c)}{\partial c} - \nabla  \cdot \frac{\partial \hat{\psi} (\mathbf{F}, \nabla \mathbf{F},  c,  \nabla c)}{\partial \nabla c}.
\end{eqnarray}

\noindent We note that the form of the chemical potential in Eq.~(\ref{53}), derived from a microforce balance and thermodynamically consistent constitutive relations, is the same as the chemical potential obtained from a variational derivative of the energy functional \citep{rudraraju2016mechanochemical, zhang2018sodium}.

\vspace{2mm}
\noindent Finally, we write the constitutive form of the flux of the diffusing species $\mathbf{j}$ in Eq.~(\ref{48}) as:
\begin{eqnarray} \label{54}
\mathbf j = -\hat{\mathbf{M}} (\mathbf{F}, \nabla \mathbf{F},  c,  \nabla c) \nabla \mu.
\end{eqnarray}

\noindent in which $\hat{\mathbf{M}}$ is a mobility tensor. Substituting Eq.~(\ref{54}) in Eq.~(\ref{52}), we express the dissipation inequality as:
\begin{eqnarray} \label{55}
\nabla  \mu  \cdot \hat{\mathbf{M}} (\mathbf{F}, \nabla \mathbf{F},  c,  \nabla c)  \nabla  \mu \geq 0.
\end{eqnarray}

\noindent The inequality in Eq.~(\ref{55}) requires the mobility tensor $\hat{\mathbf{M}}$ to be a positive-semidefinite quantity.

\section{Constitutive Theory for Intercalation Materials} \label{sec3}

\noindent The theory presented in section~\ref{sec2} is general and applicable to all first-order phase transformation materials, undergoing a symmetry-lowering lattice transformation. We next adapt this theory for intercalation materials, by introducing special constitutive equations that are relevant for modeling the Cahn-Hilliard type of diffusion and couple it with the structural transformations of lattices.

\subsection{Free Energy Density} 

\noindent We now propose a free energy density for intercalation materials undergoing a symmetry-lowering lattice transformation. This energy density is applicable to other chemo-mechanically coupled phase transformation materials undergoing a displace-type of transformation. 

\vspace{2mm}
\noindent Let us begin with the free energy density $\hat{\psi} (\mathbf{F}, \nabla \mathbf{F},  c,  \nabla c)$ that depends on the lattice deformation gradient $\mathbf{F}$, $\nabla \mathbf{F}$ and the concentration of diffusing species $c$, $\nabla c$. We assume that this free energy density is defined and continuous for all $\mathbf{F} \in \mathrm{M}^{3 \times 3}$ and $\mathrm{det}\mathbf{F} > 0$ and for all values of the species concentration $c$ during composition evolution. Here $\mathrm{M}^{3 \times 3}$ denotes a set of real $m \times n$ matrices. We assume that the free energy is Galilean invariant: for all $\mathbf{F} \in \mathrm{M}^{3 \times 3}~\mathrm{and}~\mathrm{det}\mathbf{F} > 0$, for all values of $c$ around the critical point, and each orthogonal rotations $\mathbf{R}$ with $\mathrm{det}\mathbf{R} = 1$, we have:
\begin{eqnarray}\label{55a}
    \hat{\psi} (\mathbf{RF}, \nabla (\mathbf{RF}),  c,  \mathbf{R}\nabla c)=\hat{\psi} (\mathbf{F}, \nabla \mathbf{F},  c,  \nabla c).
\end{eqnarray}

\noindent Or equivalently, we describe the free energy in terms of the symmetric Green-Lagrange strain tensor from Eq.~(\ref{5.42}) $\mathbf{E} = \frac{1}{2}(\mathbf{F}^{\top}\mathbf{F} - \mathbf{I})$, such that:
\begin{eqnarray}
    \hat{\psi} (\mathbf{F}, \nabla \mathbf{F},  c,  \nabla c)=\psi (\mathbf{E}, \nabla \mathbf{E},  c,  \nabla c).
\end{eqnarray}

\noindent Furthermore, we assume that the free energy density is an isotropic function of its arguments, and consequently depends only on the magnitude of the gradient terms $\nabla\mathbf{E}$ and $\nabla {c}$,
\begin{eqnarray}
    \psi (\mathbf{E}, \nabla \mathbf{E},  c,  \nabla c)=\psi(\mathbf{E},\vert\nabla\mathbf{E}\vert,c,\vert\nabla c\vert).
    \label{eq:}
\end{eqnarray}

\noindent We note that this free energy density satisfies frame-indifference and material symmetry, i,e., for all rotations $\mathbf{R} \in \mathcal{P}(\mathbf{e}^\circ_i)$ and $\mathcal{P}(\mathbf{e}^\circ_i)$ is a finite point group of the undistorted crystalline lattice. The free energy density $\psi$ therefore satisfies:
\begin{eqnarray}\label{eq:lag strain}
    \psi(\mathbf{RER}^{\top},|\nabla\mathbf{E}|,c,|\nabla c|) = \psi(\mathbf{E},\vert\nabla\mathbf{E}\vert,c,\vert\nabla c\vert)\quad \forall~\mathbf{R} \in \mathcal{P}(\mathbf{e}^\circ_i).
\end{eqnarray}

\noindent The free energy density in Eq.~(\ref{eq:lag strain}) is a higher-order polynomial of its arguments and has a multi-well landscape. This together with the the higher-rank tensors in Eq.~(\ref{eq:lag strain}) introduces nonlinearities making it a challenge to numerically solve the free energy functional. We overcome some of these challenges, by writing the free energy density in terms of a symmetry-adapted strain measure vector $\mathbf e = (e_1, e_2, e_3, e_4, e_5, e_6)^{\top}$:
\begin{eqnarray} \label{56}
{\psi}(\mathbf{E}, \nabla \mathbf{E},  c, \nabla c)=
{\psi} (\mathbf e, \nabla \mathbf e,  c, \nabla c).
\end{eqnarray}

\noindent Following \citep{barsch1984twin,thomas2017exploration}, we use $\mathbf{e}$ to describe the symmetry-breaking structural transformations of the lattices. The components of this strain measure are in turn a linear combination of the Green-Lagrange strain tensor components. That is, the strain measures $\mathbf{e} = \{e_1, e_2, \dots, e_6\}^T$ are defined using the Green-Lagrange strain tensor components in three dimensions as follows:
\begin{eqnarray} \label{57}
e_1 &=& \frac{1}{\sqrt {3}}(E_{11}+E_{22}+E_{33}),\nonumber \\ 
e_2 &=& \frac{1}{\sqrt {2}}(E_{11}-E_{22}),\nonumber \\ 
e_3 &=& \frac{1}{\sqrt {6}}(E_{11}+E_{22}-2E_{33}),\nonumber \\ 
e_4 &=& \sqrt {2}E_{23} = \sqrt {2}E_{32},\nonumber \\ 
e_5 &=& \sqrt {2}E_{13} = \sqrt {2}E_{31},\nonumber \\ 
e_6 &=& \sqrt {2}E_{12} = \sqrt {2}E_{21}.
\end{eqnarray}

\noindent We note that within the limits of a small deformation framework, $e_1$ represents a stretch-like deformation, and $e_4$, $e_5$ and $e_6$ describe the shear-like deformation of lattices. The strain measures $e_2$ and $e_3$ in Eq.~(\ref{57}) not only assume different values in relation to a high-symmetry cubic lattice and a lower-symmetry tetragonal lattice, but also assume separate values to differentiate among the three tetragonal lattice variants. This construction of $e_2$ and $e_3$, therefore, is most suitable as structural order parameters describing the cubic-to-tetragonal lattice transformations in intercalation materials.

\vspace{2mm}
\noindent It follows from Eqs.~(\ref{eq:lag strain}) and (\ref{56}) that the free energy density satisfying both frame-indifference and material-symmetry is
\begin{eqnarray}\label{55d2}
\psi(\mathbf{Re},\mathbf{R}\nabla(\mathbf{e})\mathbf{R}^{\top},c,\mathbf{R}\nabla c) = \psi(\mathbf{e},\nabla\mathbf{e},c,\nabla c)
\end{eqnarray}

\noindent and holds for all $\mathbf{R} \in \mathcal{P}(\mathbf{e}^\circ_i)$. In this work, we assume a zero vector $\mathbf{e} = 0$ minimizes the free energy density in the reference state and a non-zero tensor $\mathbf{e} \neq 0$ minimizes the free energy in the intercalated state. The invariant condition in Eq.~(\ref{55d2}) implies that if a pair $(\mathbf{e},c)$ minimizes $\psi$, so is $(\mathbf{Re}, c)$ for each $\mathbf{R} \in \mathcal{P}(\mathbf{e}^\circ_i)$. These minimizers correspond to the different variants of the lower-symmetry phase and contribute to the multi-well energy landscape of the free energy function.

\vspace{2mm}
\noindent We next prescribe the specific form of the free energy function. We start by decomposing the free energy density into bulk and gradient energy terms. The bulk energy, in turn, includes contributions from the thermodynamic, elastic, and chemo-mechanically coupled terms:
\begin{eqnarray} \label{58}
\psi(\mathbf e, \nabla \mathbf e,  c, \nabla c) &=&  \psi_{\mathrm{bulk}}(\mathbf{e},c) +  \psi_{\mathrm{grad}} ({\mathbf e}, \nabla \mathbf e,  c, \nabla c) \nonumber \\
&=& \psi_{\mathrm{ther}} (c)+\psi_{\mathrm{elas}} (\mathbf e)+\psi_{\mathrm{coup}} (\mathbf e, c)+\psi_{\mathrm{grad}} (\mathbf e, \nabla \mathbf e, c, \nabla c).
\end{eqnarray}

\vspace{2mm}
\noindent We construct the thermodynamic energy contribution $\psi_{\mathrm{ther}}(c)$ as a sum of an ideal solution entropy and an excess free energy---representing a deviation from thermodynamic ideality---using a Redlich-Kirster polynomial series \citep{redlich1948activity}:

\begin{eqnarray} \label{eq4} 
\psi_{\mathrm{ther}}(\bar c) &=& RT_0c_{0}\left(\frac{T}{T_0}\left[\bar{c}\operatorname{ln} \bar{c}+\left(1-\bar{c}\right)\operatorname{ln}\left(1-\bar{c}\right)\right]+\mu_0 \bar{c}+\bar{c}(1-\bar{c})\sum_{i=1}^{n}\alpha_{i}(1-2\bar{c})^{i-1}\right).
\end{eqnarray}

\noindent Eq.~(\ref{eq4}) describes an energy landscape, which distinguishes between the reference and intercalated phases, as a function of the normalized concentration $\bar c$ (scaled with the maximum species concentration $c_0$ as $\bar c=c/c_0$). The constants, $R$ and $T_0$ correspond to the gas constant and the reference temperature, respectively. The Redlich-Kirster coefficients $\alpha_{i}$ are obtained from the least square method and represent the excess energy contribution.

\vspace{2mm}
\noindent The elastic energy term penalizes the bulk and shear deformations:
\begin{eqnarray}  
 \psi_{\mathrm{elas}}(\mathbf{e})=\mathit{K}(e_1-\Delta V (e^2_2+e^2_3))^2 + \mathit{G}(e^2_4+e^2_5+e^2_6).
\end{eqnarray}

\noindent The coefficients $\mathit{K}$ and  $\mathit{G}$ correspond to the bulk and shear modulus, respectively, and $\Delta V$ represents the volume change associated with the cubic to tetragonal structural transformation of the host lattices.

\vspace{2mm}
\noindent Following \cite{shchyglo2012martensitic,barsch1984twin,ahluwalia2006dynamic}, we next construct the coupled chemo-mechanical energy term $\psi_{\mathrm{coup}}(\mathbf{e},\bar c)$ that describes a multi-well energy landscape in terms of order parameters $\bar{c}, e_2, e_3$. This energy landscape governs the cubic to tetragonal lattice transformation:
\begin{eqnarray} \label{eq90} 
\psi_{\mathrm{coup}}(\mathbf e, \bar{c}) = \beta_1(\bar{c}) (e^2_2+e^2_3) + \beta_2(\bar{c}) e_3 (e^2_3-3e^2_2) + \beta_3(e^2_2+e^2_3)^2.
\end{eqnarray}

\begin{figure}[ht!]
    \centering
    \includegraphics[width=0.5\textwidth]{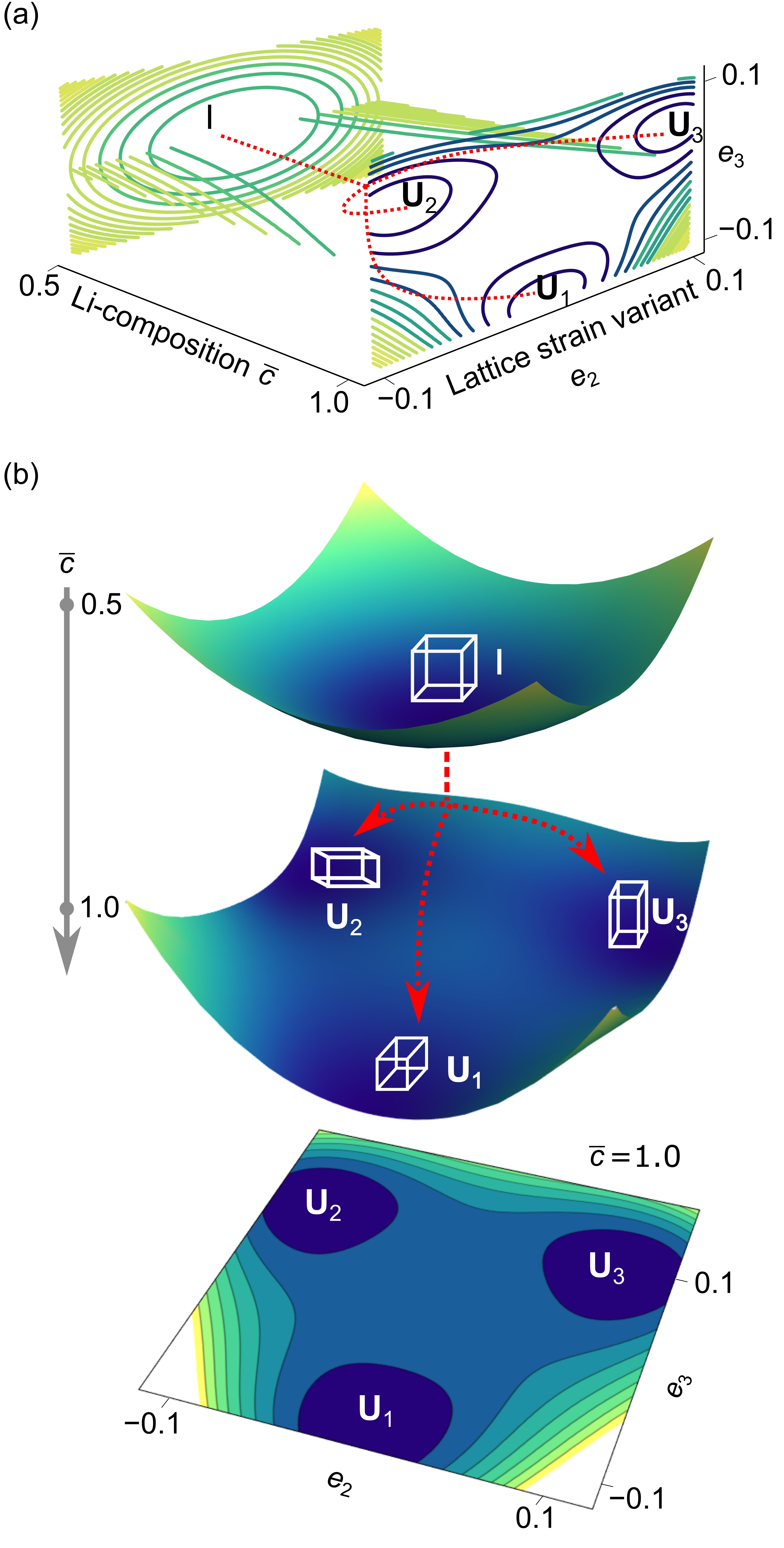}
    \caption{(a) A contour plot of the free energy function with energy wells located at $\bar{c} = 0.5$ and $\bar{c} = 1.0$. At $\bar{c} = 1.0$, three energetically equivalent wells, corresponding to the three tetragonal lattice variants are described. (b) An alternative representation of the multiwell energy landscape. The energy landscape with a single well at $\bar{c} = 0.5$ (corresponding to a high-symmetry cubic phase) transforms to a multiwell energy landscape at $\bar{c} = 1.0$. The red-dashed lines illustrate the energy-minimizing pathways between the cubic and tetragonal phases.}
    \label{Fig4}
\end{figure}

\noindent The coefficient $\beta_1(\bar{c})$ represents the concentration-dependent deviatoric modulus governing the cubic to tetragonal transformations. The coefficients $\beta_2$ and $\beta_3$ are the nonlinear elastic constants. The third-order terms accompanying concentration-dependent $\beta_2$ coefficient are required to describe a first-order type of phase-transformation \citep{cowley1976acoustic}. Fig.~\ref{Fig4}(a-b) shows three-dimensional plots of the free energy as a function of strain measure components $e_2$, $e_3$ and normalized concentration $\bar c$. The energy well at $(e_2,e_3,\bar{c}) = (0,0,0.5)$ corresponds to the higher-symmetry cubic phase (denoted by Identity tensor $\mathbf{I}$) and the energy wells at $(e_2,e_3,\bar{c}) \in \{(0,-0.1,1.0),(-0.1,0.1,1.0),(0.1,0.1,1.0)\}$ corresponds to the tetragonal variants (denoted by stretch tensors $\mathbf{U}_1, \mathbf{U}_2, \mathbf{U}_3$) at the lower-symmetry phase.

\vspace{2mm}
\noindent Finally, the gradient energy term penalizes changes in concentration and/or strain and is given by: 
\begin{eqnarray} \label{59}
\psi_{\mathrm{grad}}(\mathbf e, \nabla \mathbf e, \bar c, \nabla \bar{c})
&=& \frac{{RT_0c_0}}{2} \left(\nabla \bar c \cdot \boldsymbol{\lambda} (\bar c, \mathbf e) \nabla \bar c +\sum_{i,j} \nabla e_{i} \cdot \boldsymbol{\kappa}^{ij} (\bar c, \mathbf e) \nabla  e_{j} \right.\nonumber\\
&& \left.+2\sum_{i} \nabla \bar c \cdot  \boldsymbol{\gamma}^{i} (\bar c, \mathbf e)  \nabla e_{i}\right).
\end{eqnarray}

\noindent In Eq.~(\ref{59}) we only include energy contributions from the quadratic gradient terms that also account for the mixed terms between $\nabla\bar c$ and $\nabla\mathbf e$. These gradient terms describe nonlocal elastic and composition behavior, and their corresponding coefficients, in their general forms, are tensors that can be functions of both concentration and strain. Specifically, the concentration gradient energy coefficient is denoted by $\boldsymbol{\lambda}$ a symmetric tensor, the strain gradient energy coefficients are denoted by tensors $\boldsymbol{\kappa}^{ij}$ for each combination of $i,j =1,\dots,6$, and mixed gradient energy coefficient(s) $\boldsymbol{\gamma}^{i}$ is a tensor for $i = 1,\dots, 6$. Eqs.~(\ref{eq4})-(\ref{59}) collectively construct the total free energy density.

\subsection{Stress, Chemical Potential, and Microscopic Force}
\noindent We next derive the constitutive equations for the stresses, chemical potential, and microscopic forces for the form of the free energy function in Eq.~(\ref{58}). Using Eqs.~(\ref{58}) and (\ref{51}a) the first Piola-Kirchhoff stress tensor is given by:
\begin{eqnarray} \label{60}
\mathbf{T}_{\mathrm{R}}= \sum_{i}\left(\frac{\partial \psi}{\partial e_i} \frac{\partial e_i}{\partial \mathbf{F}} + \frac{\partial \psi_{\mathrm{grad}}}{\partial \nabla e_{i}} \frac{\partial\nabla e_{i}}{\partial \mathbf{F}}\right) ~ \mathrm{with} ~ i=1,\dots,6.
\end{eqnarray}

\noindent We write the third-order stress tensor using Eqs.~(\ref{51}b) and (\ref{58}) as:
\begin{eqnarray} \label{61}
\mathbf{Y} =  \sum_{i}\frac{\partial \psi_{\mathrm{grad}}}{\partial\nabla e_{i}} \frac{\partial\nabla e_{i}}{\partial \nabla\mathbf{F}}~ \mathrm{with} ~ i=1,\dots,6.
\end{eqnarray}

\noindent We derive the chemical potential expression using Eq.~(\ref{53}) and (\ref{58}) as follows:
\begin{eqnarray} \label{eq32}
\mu
&=& \frac{1}{c_0}\frac{\partial \psi_{\mathrm{bulk}}}{\partial \bar{c}} + \frac{RT_0}{2}\left(\nabla \bar c \cdot \frac{\partial \boldsymbol{\lambda}}{\partial \bar c} \nabla \bar c + \sum_{i,j} \nabla e_{i} \cdot  \frac{\partial  \boldsymbol{\kappa}^{ij}}{\partial \bar c} \nabla e_{j} + 2\sum_{i}\nabla \bar c  \cdot \frac{\partial \boldsymbol{\gamma}^{i}}{\partial \bar c} \nabla e_{i}\right) \nonumber \\
&& - RT_0 \left( \nabla  \cdot (\boldsymbol{\lambda} \nabla \bar c) +  \sum_{i}\nabla  \cdot(\boldsymbol{\gamma}^{i} \nabla e_{i})\right).
\end{eqnarray}

\noindent Finally, we construct the vector microscopic force  using Eq.~(\ref{58}) and Eq.~(\ref{51}d):
\begin{eqnarray} \label{62}
\boldsymbol{\xi} = \boldsymbol{\lambda} \nabla \bar{c} + \sum_{i}\boldsymbol{\gamma}^i \nabla e_{i}.
\end{eqnarray}

\noindent We discretize these constitutive equations and implement them in a finite element framework in section \ref{sec4}.

\subsection{Electrochemical reaction} 

\noindent During intercalation a guest-species, such as lithium, is inserted into the host-material lattices. This insertion is accompanied by an electrochemical reaction, at the electrode/electrolyte interface, which can be described for any guest species `A' as follows:
\begin{eqnarray}
A^{+}+e^- \rightarrow A.
\end{eqnarray}

\noindent We incorporate this electrochemical reaction in our constitutive model by using a phenomenological Butler-Volmer equation \citep{bai2011suppression, mykhaylov2019elementary, ganser2019extended} and describing the reaction rate. The Butler-Volmer reaction relates the current density $i$ to surface overpotential $\eta$ at the interface between electrode and electrolyte as:
\begin{eqnarray} \label{eqbv} 
i=i_0\left[\mathrm{exp}\left(\left(1-\beta\right)\frac{F\eta}{RT_0}\right)-\mathrm{exp}\left(-\beta\frac{F\eta}{RT_0}\right)\right].
\end{eqnarray}

\noindent In Eq.~(\ref{eqbv}) $\beta$ is the electron-transfer symmetry factor, and $F$ is the Faraday constant. The exchange current density $i_0$ given by:
\begin{eqnarray} \label{eqi0} 
i_0 = k_0 F (1-\bar c) \mathrm{exp}\left(\frac{(1-\beta)\mu_+}{RT_0}\right) \mathrm{exp}\left(\frac{\beta\mu}{RT_0}\right).
\end{eqnarray}

\noindent with $k_0$ denoting the reaction rate constant (units of $\mathrm{mol/m^2 s}$) and $\mu_+$ denoting the chemical potential in the electrolyte. We assume that the guest-species moves much faster in the electrolyte than in the active host material and therefore set $\mu_+ = 0$ \citep{bai2011suppression}. We define the surface overpotential $\eta$ as
\begin{eqnarray}  \label{eqeta} 
\eta=\Delta\phi - \frac{\mu_+ - \mu}{F} = \Delta\phi + \frac{\mu}{F}.
\end{eqnarray}

\noindent where $\Delta\phi$ is the voltage drop across the interface between electrode and electrolyte. This voltage drop serves as a driving force for species insertion (or removal) into the electrode. 

\vspace{2mm}
\noindent By combining Eqs.~(\ref{eqbv})-(\ref{eqeta}) we write the species flux at the electrode/electrolyte interface as:
\begin{eqnarray} \label{J} 
j_{n}=-\frac{i}{F}=k_0  (1-\bar c)  \mathrm{exp}\left(\frac{\beta\mu}{RT_0}\right)\left[\mathrm{exp}\left(-\beta\frac{F\eta}{RT_0}\right)- \mathrm{exp}\left(\left(1-\beta\right)\frac{F\eta}{RT_0}\right) \right].
\end{eqnarray}

\noindent Eq.~(\ref{J}) is important in modeling the galvanostatic boundary conditions for a battery electrode. For example, in this work we model the galvanostatic charge and discharge boundary conditions by employing the Butler-Volmer equation Eq.~(\ref{J}) on the electrode surface. On all reactive boundaries $\partial \Omega^{\{\mathbf{j}\}}$ we model a global flux $I$ [mol/s] by summing over each boundary $\partial \Omega^{\{\mathbf{j}\}_k}$ as follows: 

\begin{eqnarray}  \label{eq69}
    I = \sum_k\int_{\partial \Omega^{\{\mathbf{j}\}_k}} \mathbf j  \cdot \hat{\mathbf {n}}dA= \sum_k\int_{\partial \Omega^{\{\mathbf{j}\}_k}} j_n  dA = \frac{c_0\mathcal{C}}{3600}\int_{\Omega}dV.
\end{eqnarray}

\noindent In Eq.~(\ref{eq69}) the C-rate $\mathcal{C}$ is defined as the rate of time (in hours) required to charge or discharge a battery and is divided by 3600 to keep all time-related units in seconds. We use Eqs.~(\ref{eqeta})-(\ref{eq69}) to compute the voltage drop $\Delta\phi$ across the electrode/electrolyte surface as described in \ref{Appendix B}.

\vspace{2mm}
\noindent Finally, we introduce Damk\"ohler number $\mathrm{Da}$ to compare the reaction and diffusion time scales:
\begin{eqnarray} 
\mathrm{Da} = \frac{L k_0}{D_{0}c_0}.
\end{eqnarray}

\noindent where $L$ is the characteristic length scale in the model, and $D_0$ is the diffusion coefficient. For a more detailed description of the galvanostatic (dis-)charging using the Butler-Volmer equation, we refer the reader to \ref{Appendix B}.

\subsection{Governing Equations and Boundary Conditions}\label{sec:3.4}

\noindent In this section we summarize the governing equations and boundary conditions for an intercalation material undergoing a first-order symmetry-lowering phase transformation:

\begin{itemize}
    \item \textit{Concentration Evolution}: By substituting the constitutive form of flux of the diffusing species $\mathbf{j}$ in Eq.~(\ref{54}) into the local mass balance law for the species concentration in Eq.~(\ref{13}), we have:
    \begin{eqnarray} \label{eq33}
    \frac{\partial c}{\partial t} ={\hat{\mathbf{M}}} (\mathbf{F}, \nabla \mathbf{F},  c,  \nabla c)  \nabla  \mu.
    \end{eqnarray}

    Please note that the  chemical potential $\mu$ in Eq.~(\ref{eq33}) is defined in Eq.~(\ref{eq32}).

    \item \textit{Macroscopic Force Balance}: We use the local force balance law in Eq.~(\ref{28}):
    \begin{eqnarray} \label{64}
        \nabla \cdot \mathbf{T}_{\mathrm{R}}^{\top}-\nabla \cdot(\nabla \cdot \mathbf{Y}^{\top})^{\top}+\mathbf{b}=0.
    \end{eqnarray}
    
    In Eq.~(\ref{64}) $\mathbf{b}$ represents the non-inertial body force, and the stresses $\mathbf{T}_{\mathrm{R}}$ and $\mathbf{Y}$ are given by Eq.~(\ref{60}) and Eq.~(\ref{61}), respectively. Please note that the higher-order stresses $\mathbf{Y}$ are often absent in traditional elasticity problems, however, we include these stresses in our work making the local force balance a fourth-order nonlinear PDE. This introduces additional challenges in solving the numerics that we discuss in the next section.

    \item \textit{Mechanical Boundary Conditions}: To define the mechanical boundary conditions on the intercalation material, we consider $\partial \Omega = \partial \Omega^{\{\boldsymbol{\chi}\}} \cup \partial \Omega^{\{\mathbf{t}\}}$ are complementary subsurfaces of the boundary, in which the motion $\boldsymbol{\chi}$ is specified on $\partial \Omega^{\{\boldsymbol{\chi}\}}$ and the surface traction $\mathbf{t}$ on $\partial \Omega^{\{\mathbf{t}\}}$.
    $\zeta^L$ denotes a smooth boundary edge on which a jump in higher-order
    stress traction may occur \citep{toupin1962elastic}. Considering the symmetry condition of $\mathbf{Y}$, boundary conditions are finally given by:
    \begin{eqnarray} \label{66}
    \boldsymbol{\chi}&=&\boldsymbol{\breve{\chi}} ~ \mbox{on}~ \partial \Omega^{\{\boldsymbol{\chi}\}},\nonumber \\  
    \mathbf{T}_{\mathrm{R}}\hat{\mathbf{n}}-(\nabla\mathbf{Y}\cdot\hat{\mathbf{n}})\colon(\hat{\mathbf{n}} \otimes \hat{\mathbf{n}})-2(\nabla^s\cdot (\mathbf{Y}^{\top})^{\top})^{\top}\hat{\mathbf{n}}-\mathbf{Y}\colon\nabla^s\hat{\mathbf{n}}&&\nonumber\\
    - \mathbf{Y}\colon ((\nabla^s \cdot \hat{\mathbf{n}})\hat{\mathbf{n}} \otimes \hat{\mathbf{n}}-\nabla^s \hat{\mathbf{n}}) &=& \mathbf{t}   ~ \mbox{on}~\partial \Omega^{\{\mathbf{t}\}},\nonumber \\ 
    \mathbf{Y}\colon(\hat{\mathbf{n}} \otimes \hat{\mathbf{n}}) &=&\mathbf{m} ~ \mbox{on}~  \partial \Omega,\nonumber \\ 
    \left[\left[\hat{n}^{\mathit{\Gamma}}_J \hat{n}_K Y_{iJK}  \right] \right]&=&l_i=0 ~ \mbox{on} ~\zeta^{L_i}.
    \end{eqnarray}
    
    The Dirichlet boundary condition in Eq.~(\ref{66}a) is of the same form that appears in classical non-gradient elasticity problems. However, its complementary Neumann boundary condition described in Eq.~(\ref{66}b) contains higher-rank tensors that introduce complexity in gradient-type elasticity problems (e.g., in the present work). In addition to Eqs.~(\ref{66}a) and (\ref{66}b), we follow Toupin's theory \citep{toupin1962elastic} and introduce a higher-order Neumann boundary condition for the higher-order stress traction in Eq.~(\ref{66}c). This form of stress traction does not have a boundary mechanism to impose a generalized moment across atomic bonds, and it is therefore defined on surface $\partial \Omega$. Finally, Eq.~(\ref{66}d) ensures that there exists no discontinuity of higher-order stress traction across a smooth boundary edge $\zeta^L$ in the absence of a balancing line traction \citep{toupin1962elastic}. We model these boundary conditions in the finite element framework as described in section~\ref{sec4}.

    \item \textit{Diffusion Boundary Conditions}: Similar to the case of describing mechanical boundary conditions, we next consider $\partial \Omega = \partial \Omega^{\{c\}} \cup \partial \Omega^{\{\mathbf{j}\}}$ are complementary subsurfaces of the boundary, in which the species concentration is specified on $\partial \Omega^{\{c\}}$ and the global flux on $\partial \Omega^{\{\mathbf{j}\}}$:
    \begin{eqnarray} \label{67}
    c&=&\breve{c} ~ \mbox{on}~  \partial \Omega^{\{c\}},\nonumber \\  
    I&=& \breve{I} ~ \mbox{on}~ \partial \Omega^{\{\mathbf{j}\}}.
    \end{eqnarray}

    Next, we note that the microscopic stresses contribute to a power expenditure by the material that is in contact with the body. This requires that we consider suitable boundary conditions on $\partial \Omega$ that involve the microscopic tractions $\boldsymbol{\xi} \cdot \hat{\mathbf n}$ and the rate of change of the species concentration $\dot {c}$. In this work, we restrict to boundary conditions resulting in a null power expenditure:
    \begin{eqnarray} \label{69}
    (\boldsymbol{\xi} \cdot \hat{\mathbf n} ) \dot c =0.
    \end{eqnarray}

    A simple boundary condition which satisfies this null  expenditure of microscopic power is given by:
    \begin{eqnarray} \label{70}
    (\boldsymbol{\lambda} \nabla c + \sum_{i}\boldsymbol{\gamma}^{i} \nabla e_{i})  \cdot \hat{\mathbf n} =0.
    \end{eqnarray}

    \item \textit{Initial Conditions}: The initial conditions are
    \begin{eqnarray} \label{71}
    \boldsymbol{\chi} (\mathbf x, 0)=\boldsymbol{\chi}_0(\mathbf x) ~ \mbox{and}~  c ({\mathbf x}, 0)= c(\mathbf x)  ~ \mbox{on}~ \Omega.
    \end{eqnarray}
    
    The coupled set of Eqs.~ (\ref{eq33})-(\ref{67}),  (\ref{70}),  and (\ref{71}) yield a initial/boundary-value problem for the motion $\mathbf \chi(\mathbf x, t)$ and the species concentration $c(\mathbf x, t)$.
    
\end{itemize}

\section{Numerical Implementation}\label{sec4}
\noindent We next outline the finite element implementation of diffusion and finite deformation theory for intercalation materials. We note that this is a coupled, nonlinear, initial boundary value problem, in which, both the Cahn-Hilliard type of diffusion and nonlinear gradient elasticity formulation are solved. These formulations include fourth-order spatial derivatives and higher-order boundary conditions, which make the computation cumbersome. For example, the numerical solutions to the fourth-order partial differential equations (Eqs.~(\ref{eq33}) and (\ref{64})) require $C^1$-continuous finite elements and, the standard $C^0$-continuous Lagrange basis functions are not sufficient. In order to reduce these continuity requirements, we follow \cite{shu1999finite}, and develop a mixed-type finite element formulation using Lagrange multipliers. In our framework, we introduce deformation gradient and chemical potential as additional degrees of freedom, and use mixed-methods to numerically solve the higher-order diffusion and nonlinear strain gradient elasticity problem.

\subsection{Macroscopic Force Balance}
\noindent Recall that the local force balance on Eq.~(\ref{64}) in the absence of non-inertial body force $\mathbf{b}$ is
\begin{eqnarray}
      \nabla \cdot \mathbf{T}_{\mathrm{R}}^{\top}-\nabla \cdot(\nabla \cdot \mathbf{Y}^{\top})^{\top}=0.
\end{eqnarray}

\noindent We introduce the deformation gradient $\mathbf{F}$ as an additional degree of freedom and enforce kinematic constraints using a Lagrange multiplier $\boldsymbol{\rho}$:
\begin{eqnarray} \label{e46}
\boldsymbol{\rho}-\nabla \cdot \mathbf{Y}^{\top}=0,\nonumber\\
\nabla \cdot \mathbf{T}_{\mathrm{R}}^{\top}-\nabla \cdot \boldsymbol{\rho}^{\top}=0.
\end{eqnarray}

% \begin{eqnarray} \label{e46}
% \boldsymbol{\rho}=\nabla \cdot \mathbf{Y}^{\top}~~\text{with}~~\rho_{iJ} = Y_{iJK,K}\nonumber\\
% \nabla \cdot \mathbf{T}_{\mathrm{R}}^{\top}-\nabla \cdot \boldsymbol{\rho}^{\top}=0~~\text{with}~~T_{R_{iJ,J}}-\rho_{iJ,J}=0
% \end{eqnarray}

\noindent The Galerkin weak form of the mixed formulation, with suitable test functions $\delta \mathbf{u}, \delta \mathbf{F}, \delta \boldsymbol{\rho}$, are given by:
\begin{eqnarray}
 \int_{\Omega} (\nabla \cdot \mathbf{T}_{\mathrm{R}}^{\top}-\nabla \cdot \boldsymbol{\rho}^{\top}) \cdot \delta \mathbf{u}  ~dV =0,
\end{eqnarray}
\begin{eqnarray}
 \int_{\Omega}(-\nabla \cdot \mathbf{Y}^{\top}+\boldsymbol{\rho}) \colon \delta \mathbf{F} ~dV =0,
\end{eqnarray}
\begin{eqnarray}
 \int_{\Omega}(\mathbf{F} - \nabla \mathbf{u} - \mathbf{I}) \colon \delta \boldsymbol{\rho} ~dV =0.
\end{eqnarray}

\noindent The corresponding forms using index notation are given by: 
\begin{eqnarray} \label{eq42}
 \int_{\Omega} (T_{R_{iJ}}-\rho_{iJ})_{,J}  \delta u_i  ~dV =0,
\end{eqnarray}
\begin{eqnarray} \label{eq43}
 \int_{\Omega}(-Y_{iJK,K} + \rho_{iJ})  \delta F_{iJ} ~dV =0,
\end{eqnarray}
\begin{eqnarray} \label{eq44}
 \int_{\Omega}(F_{iJ} - u_{i,J} - \delta_{iJ})  \delta \rho_{iJ} ~dV =0.
\end{eqnarray}

\noindent Note that these equations involve only the first-order gradients of kinematic quantities. We next introduce the boundary conditions in Eqs.~(\ref{66}b)-(\ref{66}c) to Eqs.~(\ref{eq42}) and (\ref{eq43}) and rewrite as follows:
\begin{eqnarray} \label{eq45}
\int_{\Omega}(T_{R_{iJ}}  \delta u_{i,J} -\rho_{iJ}  \delta u_{i,J})  ~dV =\int_{\partial \Omega^{\{\mathbf{t}\}}} t_i \delta u_{i} ~dA,
\end{eqnarray}
\begin{eqnarray} \label{eq46}
 \int_{\Omega}(Y_{iJK}  \delta F_{iJ,K} + \rho_{iJ} \delta F_{iJ})  ~dV =\int_{\partial \Omega} m_i \hat{n}_J \delta F_{iJ} ~dA,
\end{eqnarray}
\begin{eqnarray} \label{eq47}
 \int_{\Omega}(F_{iJ} - u_{i,J} - \delta_{iJ})  \delta \rho_{iJ} ~dV =0.
\end{eqnarray}

%  In Eq.~\textcolor{red}{\ref{}} we introduce a second-order tensor $s_{iJ}=\textcolor{red}{m_i} \hat{n}_J$ that we will use later in this work. 

\subsection{Mass Balance}

\noindent The mass balance law in Eq.~(\ref{eq33}) involves fourth-order spatial derivatives in concentration and third-order spatial derivatives in displacement, and the standard finite element method with C$^0$-continuous Lagrange basis functions are not sufficient for discretization. To overcome this numerical obstacle, we introduce the chemical potential as an additional degree of freedom and split the fourth-order PDE in Eq.~(\ref{eq33}) into two second-order equations. First, is the expression for chemical potential as given by Eq.~(\ref{eq32}) with the independent variable $c$. Second, is the concentration evolution expression described in terms of the independent variable $\mu$:
\begin{eqnarray} \label{eq48}
\frac{\partial c}{\partial t}=\nabla \cdot \left(\mathbf{M}\left(\bar c, \mathbf e\right)\nabla \mu\right).
\end{eqnarray}

\noindent We next multiply Eq.~(\ref{eq32}) and Eq.~(\ref{eq48}) with variational test functions $\delta \bar c$ and  $\delta \mu$, respectively, and integrate these equations over $\Omega$. For Eq.~(\ref{eq32}), we have:
\begin{eqnarray} 
0 &=& 
\frac{1}{c_0}\int_{\Omega}\frac{\partial \psi_{\mathrm{bulk}}}{\partial \bar{c}}  \delta \bar c~dV + RT_0\int_{\Omega}  \boldsymbol{\lambda}\nabla \bar c \cdot  \nabla (\delta \bar c) ~dV\nonumber \\ 
&&
+ \frac{RT_0}{2}\int_{\Omega} \nabla \bar c \cdot \frac{\partial \boldsymbol{\lambda}}{\partial \bar c} \nabla \bar c \delta \bar c  ~dV + \frac{RT_0}{2}\int_{\Omega}(\sum_{i,j}\nabla e_{i} \cdot  \frac{\partial  \boldsymbol{\kappa}^{ij}}{\partial \bar c} \nabla e_{j})\delta \bar c ~dV  
 \nonumber \\ 
&& + RT_0\int_{\Omega}\sum_{i} \bigl((\nabla \bar c  \cdot \frac{\partial \boldsymbol{\gamma}^{i}}{\partial \bar c} \nabla e_i)\delta\bar{c}+ \boldsymbol{\gamma}^{i} \nabla e_i\cdot  \nabla (\delta \bar c)\bigr)~dV 
 \nonumber \\
&&
- \int_{\Omega}\mu \delta \bar c~dV-RT_0\int_{\partial \Omega}(\boldsymbol{\lambda} \nabla \bar c + \sum_{i}\boldsymbol{\gamma}^{i} \nabla e_{i})  \cdot \hat{\mathbf n}\delta \bar c ~dA.
\end{eqnarray}

\noindent Similarly, for Eq.~(\ref{eq48}), we have: 
\begin{eqnarray}  \label{A3}
0= \int_{\Omega}\frac{\partial c}{\partial t} \delta \mu~dV+  \int_{\Omega} \mathbf{M}\left(\bar c, \mathbf e\right) \nabla \mu \cdot \nabla(\delta \mu)~dV-
\int_{\partial \Omega}(\mathbf{M}\left(\bar c, \mathbf e\right)   \nabla \mu  \cdot \hat{\mathbf{n}} )\delta \mu ~dA.
\end{eqnarray}

\subsection{Finite Element Implementation}

\noindent We implement the above weak forms in the open source finite-element, multiphysics framework MOOSE \citep{gaston2009moose}. We solve the system of nonlinear equations using the preconditioned Jacobian Free Newton Krylov (PJFNK) method. This approach does not require defining an explicit tangent matrix and therefore saves considerable computational time and storage. We use the implicit Backward-Euler method for time integration and an adaptive time-stepping approach for the relatively smooth diffusion process. %Please see \textcolor{red}{... for additional information on implementation... code on OSF}

\section{Application to Li$_{2x}$Mn$_2$O$_4$}

\noindent We next calibrate the material coefficients for Li$_{2x}$Mn$_2$O$_4$ ($0.5 <x < 1$) that undergoes a cubic-to-tetragonal lattice deformation during intercalation. For simplicity, we assume a two-dimensional form of the model with $E_{13} = E_{23} = E_{33} = 0$. This assumption in turn reduces the strain measures in Eq.~(\ref{57}) to:
\begin{eqnarray} \label{57_2}
e_1 &=& \frac{1}{\sqrt {2}}(E_{11}+E_{22}), \nonumber \\ 
e_2 &=& \frac{1}{\sqrt {2}}(E_{11}-E_{22}), \nonumber \\ 
e_6 &=& \sqrt {2}E_{12} = \sqrt {2}E_{21}.
\end{eqnarray}
with $e_4 = e_5 = 0$ and $e_3 = e_1/\sqrt{2}$. We therefore construct the free energy in 2D as a function of the $e_1, e_2, e_6$ and, furthermore, in the absence of experimental measurements of gradient energies, we assume the most basic expression for the gradient energy contribution. Specifically, we assume an isotropic form for the gradient energy coefficients $\boldsymbol{\lambda}=\lambda \mathbf{I}$ and $\boldsymbol{\kappa}^{ij} =\kappa^{ij}\mathbf{I}$, respectively. In the latter term, we only assume gradient energy contributions involving $\nabla e_2$ and set all other coefficients to zero. The coefficient accompanying the mixed composition-strain gradient term is also set to zero, $\boldsymbol{\gamma}^{i} = 0$. With these simplifications, the form of gradient energy density, with scalar constants $\lambda$ and $\kappa$, for Li$_{2x}$Mn$_2$O$_4$ reduces to Eq.~(\ref{eq:LMO free energy}c) and the entire form of the 2D free energy density at $T=T_0$ is given as:
\begin{eqnarray}
\label{eq:LMO free energy}
\psi_{\mathrm{ther}}(\bar c) &=& RT_0c_{0}\biggl(\left[\bar{c}\operatorname{ln} \bar{c}+\left(1-\bar{c}\right)\operatorname{ln}\left(1-\bar{c}\right)\right]+\mu_0 \bar{c}+\bar{c}(1-\bar{c})\sum_{i=1}^{n}\alpha_{i}(1-2\bar{c})^{i-1}\biggr),\nonumber\\
\psi_{\mathrm{elas}} (\mathbf e)+\psi_{\mathrm{coup}} (\mathbf e, \bar{c}) &=& \beta_1(\bar c)e_2^2 + \beta_3 e_2^4 + K(e_1 - \Delta V e_2^2)^2 + Ge_6^2, \nonumber\\
\psi_{\mathrm{grad}} (\nabla \bar c, \nabla e_2) &=& \frac{RT_0c_0}{2}(\nabla \bar c \cdot \lambda \nabla \bar c + \nabla e_{2} \cdot \kappa \nabla  e_{2}).
\end{eqnarray}

\begin{figure}[ht!]
    \centering
    \includegraphics[width=0.5\textwidth]{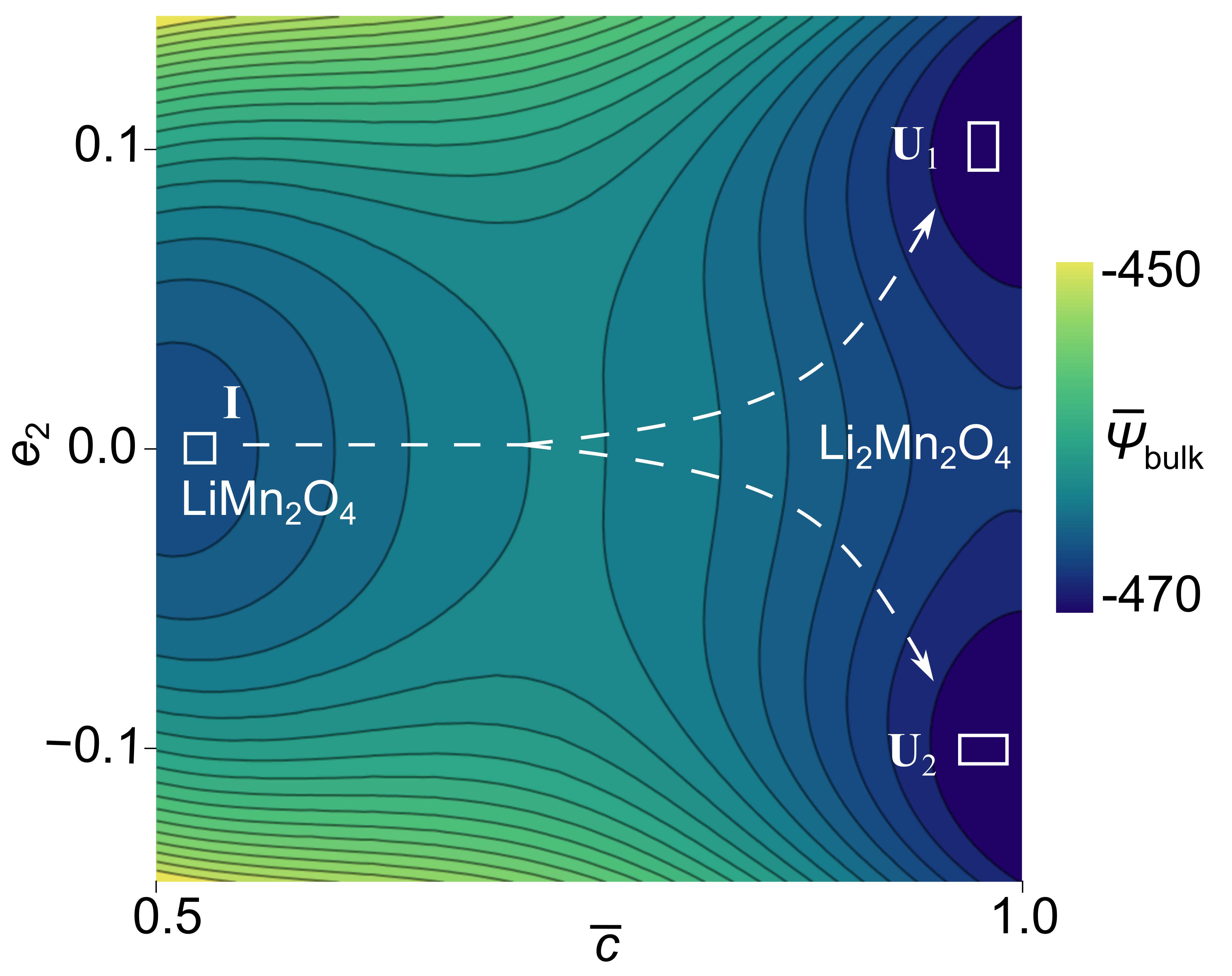}
    \caption{A multi-well energy landscape describing the symmetry-lowering structural transformation (in 2D) for Li$_{2x}$Mn$_2$O$_4$ ($0.5 <x < 1$). The higher-symmetry LiMn$_2$O$_4$ ($e_2=0$) phase is the energy minimizing deformation at $\bar{c}=0.5$. On Li-ion intercalation, LiMn$_2$O$_4$ transforms into a lower symmetry Li$_2$Mn$_2$O$_4$ phase and the lattice variants ($\mathbf{U}_1$ and $\mathbf{U}_2$) are the energy minimizing deformations at $\bar{c}=1.0$.}
    \label{Fig6}
\end{figure}

\noindent We note that in 2D, the free energy density is a functional of $\bar c$, $\nabla \bar c$, $e_1$, $e_2$, $\nabla e_2$ and $e_6$ and Fig.~\ref{Fig6} shows the multi-well energy landscape as a function of $e_2$ and $\bar{c}$. In Fig.~\ref{Fig6}, the free energy density has minima at $(\bar c,e_2) = (0.5, 0)$ corresponding to the lithium-poor phase LiMn$_2$O$_4$ and at $(\bar c, e_2) = (1.0, \pm 0.1)$ corresponding to the lithium-rich Li$_{2}$Mn$_2$O$_4$ phase, respectively. The two energy wells of equal height at $(\bar c, e_2) = (1.0, \pm 0.1)$ correspond to the two lattice variants in 2D.

\begin{figure}[ht!]
    \centering
    \includegraphics[width=0.8\textwidth]{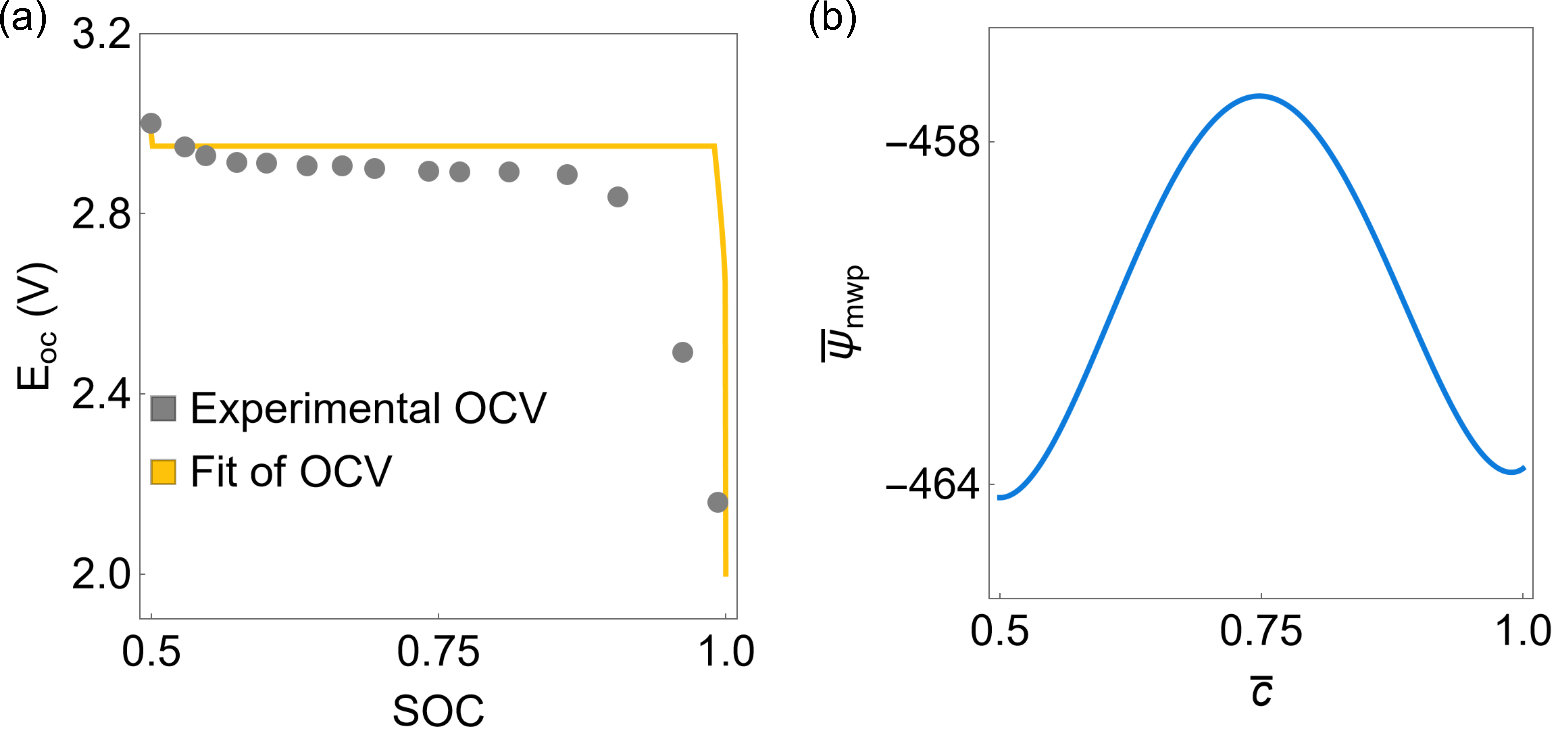}
    \caption{(a) We fit the coefficients of the thermodynamic energy term with the open circuit voltage curve measured for Li$_2$Mn$_2$O$_4$ by \citep{thackeray1983lithium}. (b) For these fitted coefficients, the normalized form of the free energy function $\bar \psi_{\mathrm{mwp}}$ is a double-well potential with minima at $\bar{c}=0.5$ and $\bar{c}=1$, respectively.}
    \label{Fig5}
\end{figure}

\vspace{2mm}
\noindent Moving forward, we nondimensionalize the total of the system as $\bar{\psi}=\psi/(RT_0c_0)$. We fit the coefficients of the thermodynamic energy term $\psi_{\mathrm{ther}}$ in Eq.~(\ref{eq:LMO free energy}a) with the phase segregation thermodynamics of Li$_{2x}$Mn$_2$O$_4$. Specifically, we fit the Open Circuit Voltage (OCV) parameters with the experimental measurements from \cite{thackeray1983lithium}, as shown in Fig. \ref{Fig5}(a). The specific values of the Redlich-Kirster and the reference chemical potential coefficients are listed in Table~\ref{T1}. For these combinations of coefficients, the Legendre transformation of the Helmholtz free energy density reduces to \citep{hormann2019phase, nadkarni2019modeling}:
\begin{eqnarray} 
\label{eq16}
\bar \psi_{\mathrm{mwp}}(\bar c) =\bar \psi_{\mathrm{ther}}(\bar c)-\frac{\partial \bar \psi_{\mathrm{ther}}(\bar c=0.5)}{\partial \bar c}\bar c
=\bar \psi_{\mathrm{ther}}(\bar c) +115.727\bar c.
\end{eqnarray}

\noindent Eq.~(\ref{eq16}) describes a doublewell structure with minima at $\bar c=0.501$ and $\bar c=0.99$, corresponding to the Li-poor (LiMn$_2$O$_4$) and Li-rich (Li$_{2}$Mn$_2$O$_4$) phases, respectively, see Fig. \ref{Fig5}(b).

\vspace{2mm}
\noindent We next calibrate the coefficients in the elastic $\psi_{\mathrm{elas}}(\mathbf{e})$ and the coupled $\psi_{\mathrm{coup}}(\mathbf{e}, \bar{c})$ energy terms for Li$_{2x}$Mn$_2$O$_4$:
\begin{eqnarray} \label{eq-elas-coup} 
\psi_{\mathrm{elas}} (\mathbf e)+\psi_{\mathrm{coup}} (\mathbf e, \bar{c}) = \beta_0\frac{\bar{c}-0.75}{0.5-0.75} e^2_2 + \beta_3 e^4_2 +K(e_1-\Delta V e^2_2)^2 + G e^2_6.
\end{eqnarray}

\noindent In Eq.~(\ref{eq-elas-coup}) the coefficients, $\beta_0 =(C_{11}-C_{12})/2$,
$K =(C_{11}+C_{12})/2$, $G =C_{44}$, are linear combinations of the elastic stiffness components $C_{11}$,  $C_{12}$, and $C_{44}$ of LiMn$_2$O$_4$. We calculate the spontaneous strains $\mathbf{E}_0$ accompanying the cubic-to-tetragonal lattice transformation of Li$_{1-2}$Mn$_2$O$_4$ and the corresponding lattice volume changes $\Delta V$ from lattice geometry measurements \citep{erichsen2020tracking}, see Table~\ref{T1}. We use these values as input to identify the energy minimizing values of the strain component $e_2 =\pm 0.1$ and calculate $\beta_3$ by solving the equilibrium equation:
\begin{eqnarray}  
\frac{\psi_{\mathrm{elas}}+\psi_{\mathrm{coup}}}{\partial e_2} &=& 0.
\end{eqnarray}

\noindent Furthermore, based on experimental reports of isotropic Li-diffusion in $\mathrm{Li_2Mn_2O_4}$ \citep{erichsen2020tracking}, we assume an isotropic form of the mobility expression with $D_0$ as the diffusion coefficient.
\begin{eqnarray}
\mathbf{M}\left(c\right) = \frac{D_0 c\left(c_{0}-c\right)}{RT_{0}c_{0}} \mathbf{I}.
\end{eqnarray}

\noindent We list all material constants calibrated to Li$_2$Mn$_2$O$_4$ in Table~\ref{T1} and compute microstructural evolution in Li$_2$Mn$_2$O$_4$ in the next section.

\begin{table}[ht!]
    \centering
    \caption{The material parameters for Li$_{2x}$Mn$_2$O$_4$ ($0.5 <x < 1$).}
    \renewcommand\arraystretch{1.5}
    \begin{tabular}{ll}
    \hline
    Parameters&Values\\
    \hline
    $\mu_0$&$-$579.454 \\
    $\alpha_1$&$-$926.715 \\
    $\alpha_2$&$-$927.453 \\
    $\alpha_3$&$-$470.114 \\
    $\lambda$&$7\times10^{-14}$ $(\mathrm{m}^2)$ \\
    $\kappa$&$7\times10^{-14}$ $(\mathrm{m}^2)$ \\
    $D_0$&$2\times10^{-14}$ $(\mathrm{m}^2/\mathrm{s})$ \citep{christensen2006mathematical}\\
    $c_0$&$4.58\times10^{4}$ $(\mathrm{mol}/\mathrm{m}^3)$  \citep{zhang2007numerical}\\
    $\mathbf{E}_0$&$\begin{pmatrix} -0.0305089& 0\\ 0& 0.130085
      \end{pmatrix}$\\
    $C_{11}$&190.75 (GPa) \citep{ramogayana2020lithium} \\
    $C_{12}$&36.63 (GPa)  \citep{ramogayana2020lithium}\\
    $C_{44}$&90.45 (GPa) \citep{ramogayana2020lithium}\\
    $\beta_0$&77.06  (GPa)  \\
    $K$&113.69 (GPa)  \\
    $G$&90.45 (GPa)  \\
    $\beta_3$&2935.82 (GPa) \\
    $\Delta V$&0.0540734\\
    \hline
    \end{tabular}
    \label{T1}
\end{table}

\newpage
\section{Results}\label{Results and Discussion}

\noindent We next apply our modeling framework to investigate the interplay between Li-diffusion and lattice deformation in Li$_2$Mn$_2$O$_4$. Specifically, we analyze the microstructural evolution process on a 2D-plane of a Li$_2$Mn$_2$O$_4$ electrode (a primary particle) during a galvanostatic discharge process, how this evolution affects the macroscopic voltage response of the material, and investigate the stresses evolving during intercalation. For our computations, we model a 2D domain of size $500\mathrm{nm} \times 500\mathrm{nm}$ of a single crystal Li$_{2x}$Mn$_2$O$_4$. We fix the displacements $\mathbf{u}=0$ of all the boundaries and apply galvanostatic discharge conditions Eq.~(\ref{eq69}), with a 5C-rate, on all the boundaries. All material constants used in our calculations are listed in Table~\ref{T1}, and for our purposes we note that the electron-transfer symmetry factor $\beta = 0.5$, the Damk\"ohler number $\mathrm{Da} = 5.6574\times10^{-3}$, and we define the state of charge as $\mathrm{SOC} = \int_{\Omega} \bar c dV/V$.

\subsection{Microstructure Evolution}\label{sec:Microstructure Evolution}

\begin{figure}[ht!]
    \centering
    \includegraphics[width=\textwidth]{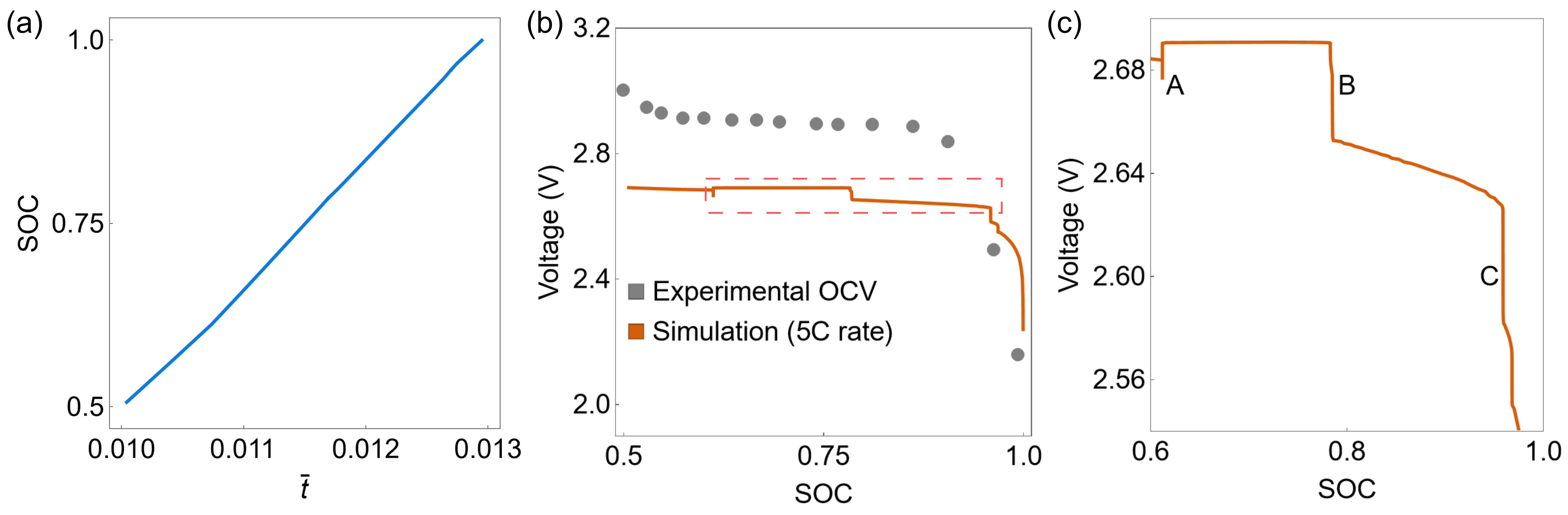}
    \caption{SOC and Voltage curves predicted by our diffusion-deformation model. (a) The SOC scales linearly as a function of the normalized time $\bar t$. (b) The Voltage curve as a function of SOC during a galvanostatic discharge at 5C-rate in our simulations. This voltage curve is lower than the experimental measurement for Li$_2$Mn$_2$O$_4$ discharge with open circuit voltage \citep{thackeray1983lithium}. (c) An inset of the voltage curve (V) showing three distinct step-like features labeled `A', `B', and `C', respectively. These steps correspond with characteristic microstructural changes that contribute to a sharp drop in the voltage.}
    \label{Fig7}
\end{figure}

\noindent During the galvanostatic discharge of Li$_{2x}$Mn$_2$O$_4$ (i.e., Li-insertion) the SOC increases linearly with time, see Fig. \ref{Fig7}(a). The corresponding voltage curve, during this discharge process at 5C-rate, plateaus at 3.0 V and is comparable with experimental measurements for Li$_{2x}$Mn$_2$O$_4$ \citep{thackeray1983lithium}. This simulated voltage curve is, however, lower than the experimental open circuit voltage measured for Li$_{1-2}$Mn$_2$O$_4$ \citep{thackeray1983lithium}. We attribute this difference between the voltage curves to the larger overpotential required to drive the galvanostatic discharge process at the 5C-rate. Additionally, we note that the voltage plateau is no longer flat but instead has a tilt/slope, which arises from the non-equilibrium operating conditions at 5C-rate. This is consistent with observations in other electrode materials such as Li$_x$CoO$_2$ \citep{nadkarni2019modeling} and Li$_x$FePO$_4$ \citep{bai2011suppression, cogswell2012coherency}. Finally, we note the appearance of step-like features marked as `A', `B', and `C' on the voltage curve in Fig.~\ref{Fig7}(b) (see inset Fig.~\ref{Fig7}(c)). These sharp step-like features correspond to the abrupt changes in the microstructural states of Li$_{2x}$Mn$_2$O$_4$ that we discuss next.

\begin{figure}[ht!]
    \centering
    \includegraphics[width=\textwidth]{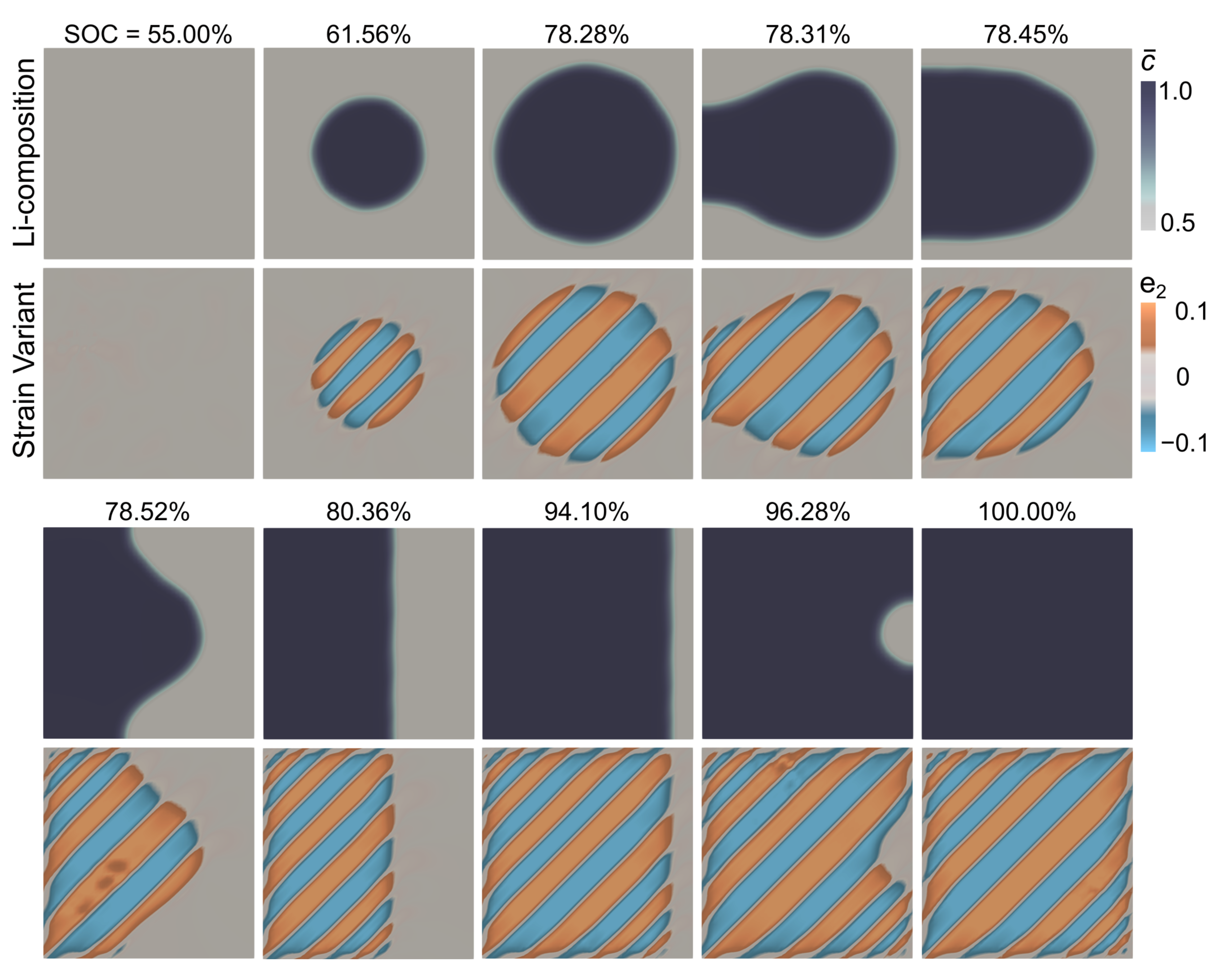}
    \caption{A representative microstructural evolution pathway predicted by our diffusion-deformation model during a galvanostatic discharge half cycle at 5C-rate. The images on the top and bottom rows, respectively, illustrate the coupled evolution of Li-composition $\bar c$ and strain $e_2$ as a function of SOC.}
    \label{Fig8}
\end{figure}

\vspace{2mm}
\noindent Fig.~\ref{Fig8} shows the microstructural evolution pathway in Li$_{1-2}$Mn$_2$O$_4$ as a function of both Li-composition $\bar {c}$ and strain $e_2$ order parameters. We initialize the computational domain with the LiMn$_2$O$_4$ phase with $\bar c = 0.5$ and the corresponding strain $e_2 = 0$. During galvanostatic discharge, the SOC of the system increases gradually and at $\mathrm{SOC} = 61.56\%$, a Li-rich phase Li$_{2}$Mn$_2$O$_4$ nucleates at the center of the domain. This change in Li-composition is accompanied by the cubic-to-tetragonal structural transformation of the host lattices, which generates two lattice variants with strains $e_2 = + 0.1$ and $e_2 = -0.1$, respectively. Each of these lattice deformations corresponds to the bottom of the energy wells at $(\bar c, e_2) = (1.0, \pm 0.1)$, however lattice misfit between the cubic-LiMn$_2$O$_4$ and the tetragonal Li$_2$Mn$_2$O$_4$ phases contributes to finite elastic energy at the phase boundary (i.e., in the interfacial region with $0.5 < \bar c < 1.0$). Minimizing this elastic energy drives the formation of twinned microstructures shown in Fig.~\ref{Fig8}.\footnote{In our algorithm, we iteratively minimize the total energy of the system using the predconditioned Jacobian Free Newton Krylov method with an adaptive time step.} That is, lattices rotate and shear to fit compatibly with each other, forming twin boundaries, and this finely twinned domain reduces the elastic energy stored at the phase boundary  \citep{ball1987fine}. 

\vspace{2mm}
\noindent The nucleation of the Li$_{2}$Mn$_2$O$_4$ phase manifests macroscopically on the voltage curve as a step-like feature `A' in Fig. \ref{Fig7}(c). This drop in the voltage curve, during a galvanostatic discharge condition, correlates with a decrease in the total free energy of the system on having overcome the nucleation energy barriers. %\textcolor{blue}{can we add as to why the voltage should drop on nucleation, something to do with overcoming nucleation energy barriers?}

\vspace{2mm}
\noindent With continued lithiation, the Li-rich Li$_{2}$Mn$_2$O$_4$ phase, grows rapidly minimizing the surfacial area of the phase boundary and further reducing the elastic energy stored in the system. At $\mathrm{SOC}=78.52\%$ the Li-rich nucleus fills the left portion of the domain and the twinned microstructures adapt to varying volume fractions, see Fig.~ \ref{Fig8}. We attribute the varying thickness of the twinned domains (i.e., the twinned domains are narrower at the domain edges in comparison to the wider twins at the domain center) to the fixed boundaries. These boundary conditions restrict deformations and the twinned microstructure adapts to minimize misfit at the domain edges. Additionally, we model zero surface wetting in Eq.~(\ref{70}), which contributes to the bending of the phase boundary at the particle surface in Fig.~\ref{Fig8}.

\vspace{2mm}
\noindent At $\mathrm{SOC}=78.52\%$, the volume fraction of the twinned domains evolves to adapt and minimize the elastic misfit at the domain boundaries. Please recall that, in our computations, we fix the displacements of the boundaries. These fixed boundaries correspond to the cubic strain variant with $e_2 = 0$ of the LiMn$_2$O$_4$ phase. On intercalation, the cubic-to-tetragonal lattice transformation generates lattice misfit at the domain boundary and the twinned microstructural pattern evolves (i.e., by changing its volume fraction) to minimize this elastic energy. This microstructural interaction is marked by the appearance of the second step-like feature `B' on the voltage curve in Fig. \ref{Fig7}(c). 

\vspace{2mm}
\noindent In the final stage of the discharge process, the phase boundary reduces to a planar geometry and propagates through the computational domain. This change in the geometry of the phase boundary is accompanied by the formation of additional twins, see Fig.~\ref{Fig8}, and the Li-composition and the twinned microstructures evolve simultaneously. At $\mathrm{SOC}=96.28\%$, the intercalation-wave interacts with the right edge of the fixed domain, and this interaction in turn corresponds to the third step-like feature `C' on the voltage curve in Fig.~\ref{Fig7}(c). At $\mathrm{SOC}=100\%$ the material is fully transformed to the Li$_2$Mn$_2$O$_4$ phase that is finely twinned, see Fig.~\ref{Fig8}. 

\begin{figure}[ht!]
    \centering
    \includegraphics[width=0.9\textwidth]{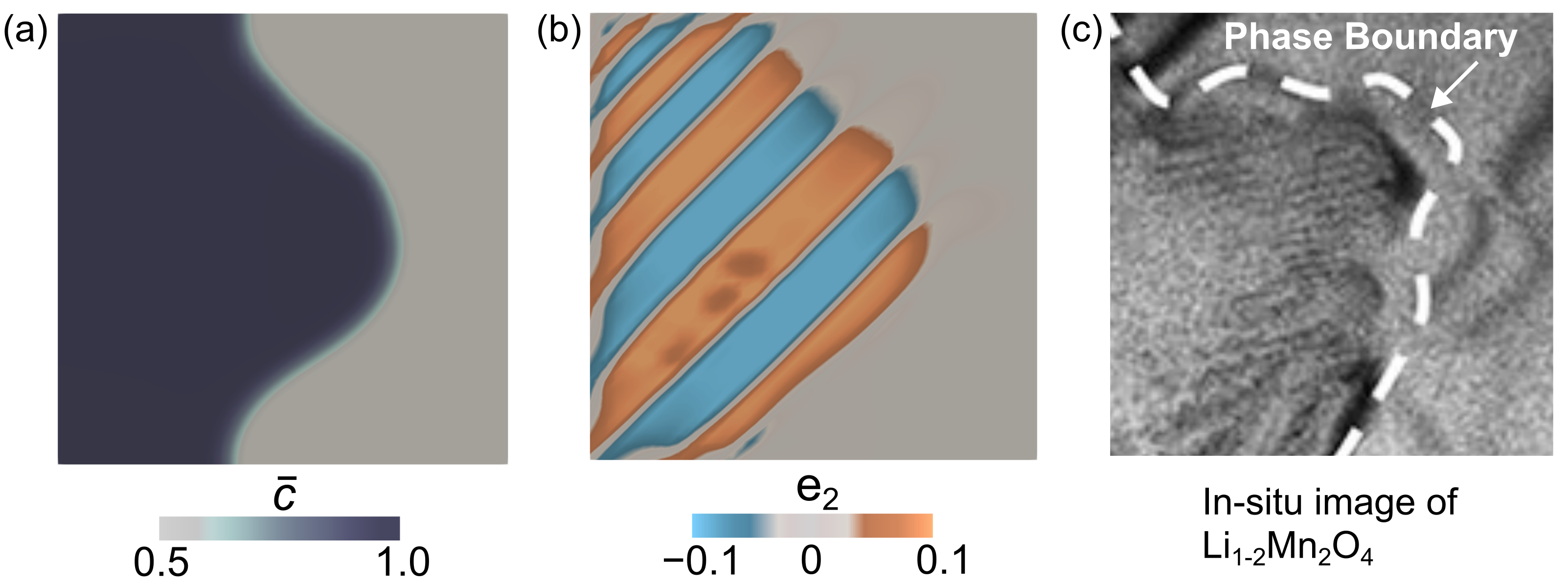}
    \caption{We compare microstructural features predicted in our simulation with the experimental image of Li$_{2x}$Mn$_2$O$_4$. (a-b) The Li-composition and strain variant distribution at SOC = 78.52\% shows a curved phase boundary and finely twinned microstructures. This prediction compares favorably with the bright field imaging Li$_{2x}$Mn$_2$O$_4$ by \citep{erichsen2020tracking}(Reproduced with permission from American Chemical Society).}
    \label{Fig9}
\end{figure}

\vspace{2mm}
\noindent We compare our modeling predictions with previously published in-situ bright field imaging of microstructural patterns in Li$_{2x}$Mn$_2$O$_4$ \citep{erichsen2020tracking}, see Fig.~\ref{Fig9}. We note similarities between our simulations and the experimental images and also highlight the differences: First, the phase boundary separating the Li$_{2}$Mn$_2$O$_4$/LiMn$_2$O$_4$ phases is curved in both our simulation and the experimentally imaged microstructure. This curved morphology does not correspond to the energy-minimizing orientation of the phase boundary in equilibrium and arises from the dynamic galvanostatic boundary conditions applied on the electrode surface. Second, the appearance of twinned domains in the Li$_{2}$Mn$_2$O$_4$ is another commonality, however, the volume fraction of the tetragonal twins differ between the experiment ($f = 0.2$, see section \ref{sec2.3} and \cite{erichsen2020tracking}) and our simulation ($f = 0.5$). We attribute this difference to the far-from-equilibrium driving conditions and the fixed boundaries modeled in our computations. That is, in Fig.~\ref{Fig9} we model a 500nm domain with fixed boundaries and a galvanostatic discharge (5C-rate) condition. This differs from the $\sim4\mu\mathrm{m}$ bulk-type free-standing electrode in the experiment \cite{erichsen2020tracking}, in which the lithium tip is in contact only at the electrode surface and experimental discharge conditions are closer to the equilibrium state. It is important to note that our analytical solutions are consistent with the geometric features of the experimentally imaged microstructure, see Fig.~\ref{Fig3}. Finally, as observed in experiments, the tetragonal lattice variants (with $e_2 = \pm 0.1$) nucleate independently in our computations and evolve to form compatible twin interfaces. Overall, the similarities between experiments and theoretical predictions show that our modeling framework captures the interplay between Li-diffusion and lattice deformations, and this model could in turn be used as a tool to crystallographically design microstructures in intercalation compounds.

\subsection{Stress Evolution}
\noindent We next investigate the evolution of stresses in Li$_{2x}$Mn$_2$O$_4$ electrodes during the galvanostatic discharge process. As highlighted in the previous section, Li-intercalation into LiMn$_2$O$_4$ induces an abrupt cubic-to-tetragonal lattice transformation. This structural transformation of lattices at the atomic level generates continuum stresses, which on repeated cycling, lead to particle cracking and eventual failure. With our newly developed micromechanical model, we not only capture the interplay between Li-diffusion and finite deformation of lattices in 2D, but we also predict the evolution of stresses, in-situ, during the discharge process.

\begin{figure}
    \centering
    \includegraphics[width=\textwidth]{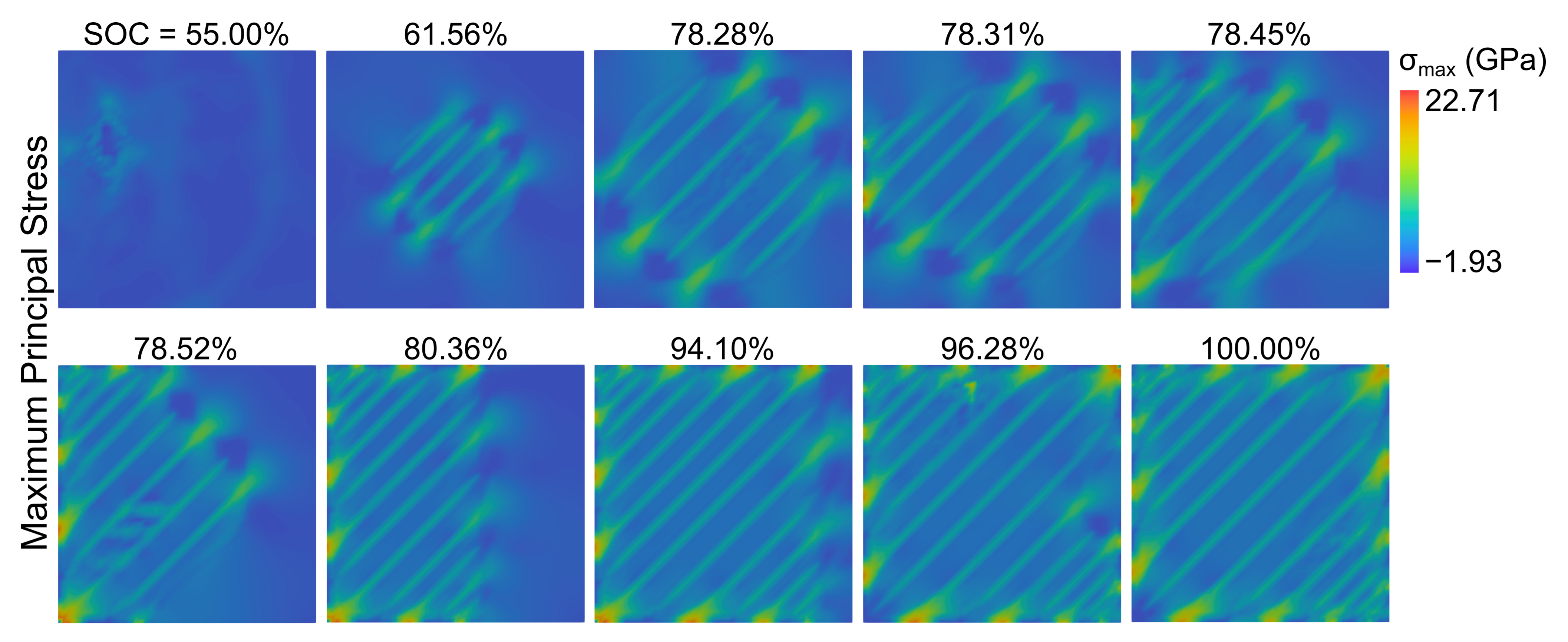}
    \caption{The maximum principal stress distribution in the computational domain corresponds to the microstructural evolution in Fig.~\ref{Fig8}. We observe stress concentrations primarily at the phase boundaries and closer to the particle surfaces. These interfacial stresses accumulate in the electrode particle with repeated cycling eventually leading to its failure.}
    \label{Fig10}
\end{figure}

\vspace{2mm}
\noindent Fig.~\ref{Fig10} shows the maximum principal stresses evolving in Li$_{2x}$Mn$_2$O$_4$ electrode during the galvanostatic discharge half cycle. This stress evolution accompanies the Li-intercalation half cycle described in Fig.~\ref{Fig8}. At the initial state, $\mathrm{SOC} = 0$, the domain is a stress-free single crystal LiMn$_2$O$_4$. On Li-intercalation, at $\mathrm{SOC} = 55\%$, compressive stresses accompanying the nucleation of the Li-rich Li$_2$Mn$_2$O$_4$ phase are observed in the electrode. It is interesting to note that tensile stresses of $\sim 4.88$ GPa are observed at the twin interfaces that separate the tetragonal lattice variants. These twin interfaces theoretically are exactly compatible interfaces that have no lattice misfit. However, in our diffuse interface modeling approach, we introduce regularization terms (i.e., gradient energy terms $\frac{1}{2}\nabla e_2 \cdot \kappa \nabla e_2$) that penalize a change in strain values. This penalty contributes to the finite stresses at the twin interface. Additionally, non-zero stresses are observed at the LiMn$_2$O$_4$/Li$_2$Mn$_2$O$_4$ phase boundary. None of the tetragonal variants of Li$_2$Mn$_2$O$_4$ fit compatibly with the cubic LiMn$_2$O$_4$ lattices. Consequently a finely twinned microstructure, comprising two variants with $e_2 = \pm 0.1$, forms to minimize the misfit strains at the LiMn$_2$O$_4$/Li$_2$Mn$_2$O$_4$. Despite the energy-minimizing deformation, finite elastic energy is stored in the phase boundary and manifests as stresses during intercalation.
%\textcolor{blue}{Thus the structural transformation proceeds coherently in the particle of Li$_{2x}$Mn$_2$O$_4$ due to the appearance of a coherent twinned microstructure.??} 

\begin{figure}[ht!]
    \centering
    \includegraphics[width=0.6\textwidth]{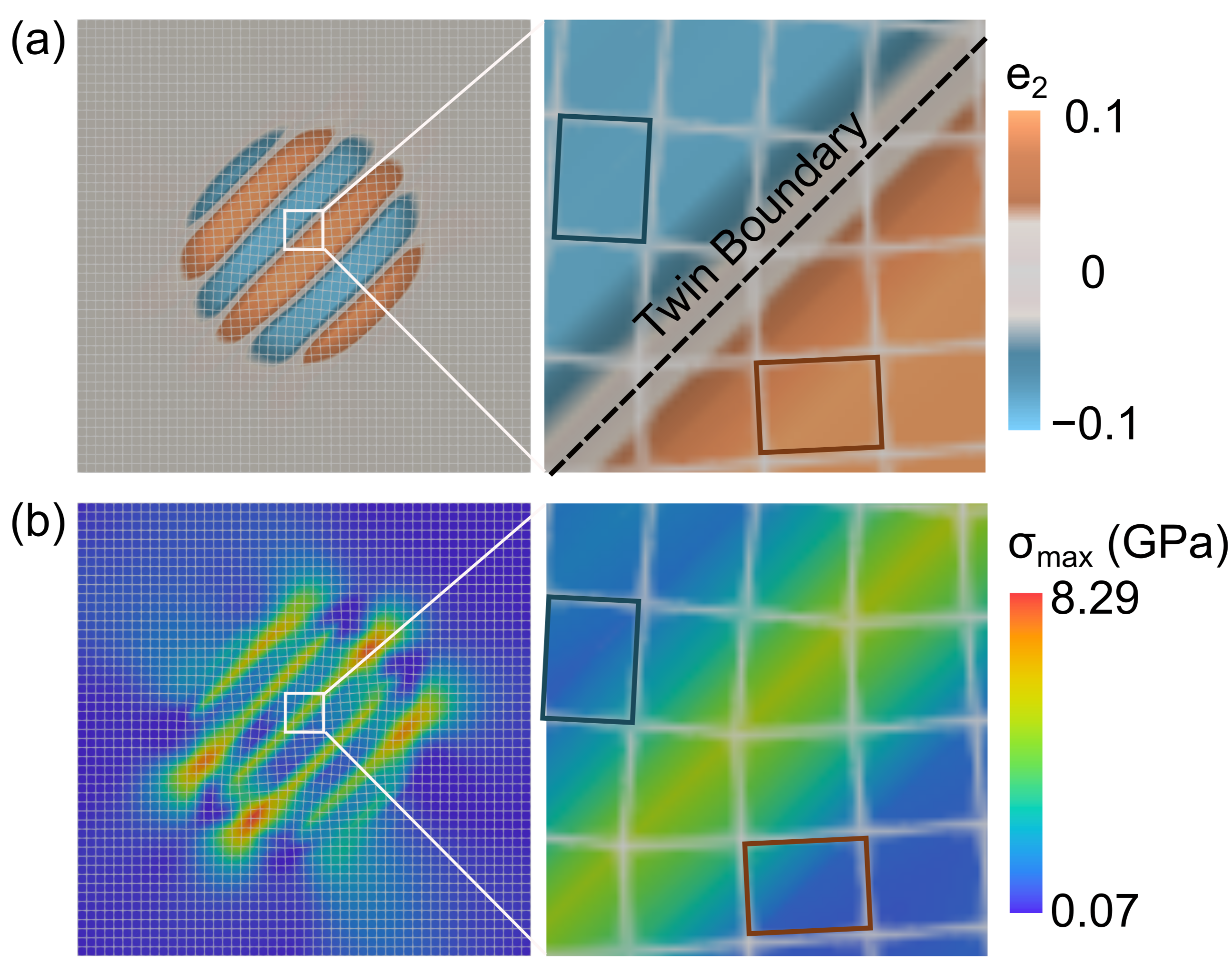}
    \caption{An inset view showing the finite deformation of the mesh during intercalation. (a) At SOC = 61.56\% a Li-rich phase (Li$_2$Mn$_2$O$_4$) nucleates and is accompanied by the cubic-to-tetragonal transformation of lattices. Two of these tetragonal variants, highlighted in the inset figure, form a compatible twin interface. (b) The maximum principal stress distribution at SOC = 61.56\% shows stress concentrations primarily at the interface separating the LiMn$_2$O$_4$/Li$_2$Mn$_2$O$_4$ phases. Tensile stresses, of a lower magnitude, are also observed along the twin boundaries and correspond to the gradient energy penalty arising from $\nabla e_2$ terms in Eq.~(\ref{eq:LMO free energy}c).}
    \label{Fig11}
\end{figure}

\vspace{2mm}
\noindent Let us take a closer look at the stress state at $\mathrm{SOC} = 61.56\%$, see Fig.~\ref{Fig11}. We highlight the cubic-to-tetragonal structural transformation of lattices in 2D using a distorted mesh grid. The undeformed square cells correspond to the Li-poor phase and the deformed rectangular cells correspond to the Li-rich phase. The different orientations of the rectangular cells represent the two lattice variants of the Li$_{2}$Mn$_2$O$_4$ phase. Note that this structural transformation is accompanied by a $5.4\%$ volume change and $\sim 13\%$ lattice strains that contribute to elastic stresses in the domain. In addition to tensile stresses along the twin boundaries, we note stress concentrations at the LiMn$_2$O$_4$ / Li$_2$Mn$_2$O$_4$ interface. These stresses correspond to the lattice mismatch between the cubic and tetragonal phases of the intercalation compound, and the tensile/compressive stresses depend on the lattice orientations at the phase boundary, see Fig. \ref{Fig11}(b).  

\vspace{2mm}
\noindent The LiMn$_2$O$_4$ / Li$_2$Mn$_2$O$_4$ phase boundary is elastically stressed, see Fig.~\ref{Fig10}, and this stressed interface moves through the electrode particle during charge/discharge processes. In experimental literature \citep{erichsen2020tracking}, this stressed interface is shown to nucleate dislocations and microcracks, that lead to eventual failure of the materials. In these experiments, the dislocations and microcrackings were observed in the proximity of the phase boundary, that is consistent with the stress distribution we observe in our simulation, see Fig.~\ref{Fig10}. Additionally, in our calculations, this stressed interface interacts with the fixed domain boundaries and contributes to increased stresses at the particle surfaces, see Fig.~\ref{Fig10} at $\mathrm{SOC} = 100\%$. These simulations provide quantitative insights into stress distributions in symmetry-lowering phase transformation materials and serve as a design tool for intercalation materials. 

%On the other hand, Fig. \ref{Fig10} highlights that the austenite-martensite interface is a semicoherent interface which has lost coherency since high stresses along the phase boundary emerge in a staggered way. This semicoherent phase boundary of Li$_{2x}$Mn$_2$O$_4$ is also experimentally verified \citep{erichsen2020tracking}. Usually, the mechanism of the loss of coherency could be related to the dislocation formation, pre-existing cracks or defects along the phase boundary \citep{cogswell2012coherency}. What is the mechanism underlying this interesting coherency loss even in an intact phase boundary? 

\section{Discussion}\label{Discussion}
\noindent We derive a thermodynamically-consistent theory that predicts the symmetry-lowering lattice transformations in first-order phase change materials. In this theory, we use the Cauchy-Born rule and the principle of virtual power to develop a multiscale modeling framework that couples finite deformation of lattices at the atomic level with the diffusion of guest species (e.g., intercalating ions) at the continuum level. We applied this theoretical framework to intercalation materials, specifically to a spinel electrode Li$_{1-2}$Mn$_2$O$_4$, and analyzed the interplay between Li-diffusion and lattice transformation during electrochemical half cycling. The theoretical predictions provide fundamental insights into microstructural evolution pathways in Li$_{1-2}$Mn$_2$O$_4$ and are consistent with the experimentally imaged HRTEM micrographs in Li$_2$Mn$_2$O$_4$ \citep{erichsen2020tracking}. Additionally, our simulations predict how individual lattice variants rotate and shear during phase transformations and how they collectively generate elastic stresses at the phase boundary. These insights indicate potential origins of structural decay in Li$_2$Mn$_2$O$_4$ (e.g., microcracking, dislocation nucleation) reported in the literature. In the remainder of this section, we discuss the limiting features of our model and then highlight its potential application for materials design.

\vspace{2mm}
\noindent Two features of our work limit the comparisons we can make with experimental observations of twinned microstructures in Li$_2$Mn$_2$O$_4$. First, we simplify the form of gradient energy contributions and evolution kinetics by assuming isotropic material constants. For example, we consider a scalar form of the gradient energy coefficients $\mathbf{\lambda} = \lambda\mathbf{I}$ and $\boldsymbol{\kappa}^{ij}= \kappa^{22}\mathbf{I} =  \kappa\mathbf{I}$ that penalizes a change in composition $\bar c$ or strain $e_2$ variable irrespective of the orientation of the interfaces. Furthermore, to prevent overfitting of the model parameters we do not penalize changes in strain components $e_1, e_6$ or the mixed terms. These gradient energy contributions could be important to describe the geometric features of twinned microstructures (e.g., orientation of the phase boundary, volume fraction of the twinned domains) and we will account for these energy terms in our future work. Despite these simplifications, our modeling predictions on twin interface orientations and the nucleation and growth of the lamellar microstructural pattern exhibit a surprisingly favorable comparison with the experimental observations.

\vspace{2mm}
\noindent Second, due to the nonlinearity and higher-order derivatives involved in the problem, we restrict ourselves to 2D finite element computations. This dimensional reduction simplifies the form of the free energy functional and we primarily describe the energy landscape using the strain variant $e_2$ as the order parameter. This 2D model phenomenologically describes the nucleation and growth of twinned domains in Li$_2$Mn$_2$O$_4$ and predicts principal stresses in 2D at the phase boundary. Individual lattices in bulk Li$_2$Mn$_2$O$_4$, however, rotate and shear in 3D space to minimize the misfit strains at the phase boundary. Computing these microstructural patterns in a 3D finite element framework is necessary to conclusively interpret microstructural evolution pathways and chemo-mechanically coupled stresses in Li$_2$Mn$_2$O$_4$. Extending our model to 3D not only presents a computational challenge, but it is also important to derive the coefficients of higher-order energy terms (e.g., nonlinear elastic energy and anisotropic gradient energy terms) using first-principle calculations \citep{zhang20233d} and/or careful experimentation. Keeping these limiting conditions in mind, we next discuss the strengths of our diffusion-deformation model and highlight its potential applications.

\vspace{2mm}
\noindent The key feature of our model is that we derive a thermodynamically-consistent diffusion-deformation theory using the virtual-power approach and the second law of thermodynamics, without specifying, apriori, the form of the free energy function. Through this approach we derive the governing equations based on classical thermodynamic arguments, which differs from other models derived using variational approaches. We numerically solve this model using mixed-type finite element methods based on Lagrange multipliers and implement our framework in an open-source MOOSE platform. Using this model, we predict the interplay between higher-order diffusion terms and nonlinear strain gradient elasticity in Li$_2$Mn$_2$O$_4$ with electrochemical boundary conditions. Unlike earlier phase field models that describe phase transformation in intercalation materials as a function of composition alone \cite{nadkarni2019modeling, zhang2019phase}, our model predicts the coupled interplay between Li-composition and finite lattice deformation and provides quantitative insights into the nucleation and growth of twinned microstructures and stress field distributions during galvanostatic cycling. These insights will play a crucial role in crystallographically designing intercalation materials and mitigating structural degradation \citep{renuka2022crystallographic, zhang2023designing, van2023ferroelastic}. More broadly, the modeling framework is applicable to describe lattice deformations in other first-order phase transformation materials, such as shape-memory alloys, multicomponent structural materials \citep{chien1998stress, krogstad2011phase} and, 2D layered nanoelectronic materials \citep{rossnagel2010suppression}.

\section{Conclusions} \label{Conclusion}
\noindent We derive a thermodynamically consistent theory that couples the diffusion of a guest species at the continuum scale with finite deformation of host lattices at the atomic scale. We adapt this diffusion-deformation theory for symmetry-lowering intercalation materials, such as Li$_2$Mn$_2$O$_4$, and predict the delicate interplay between Li-diffusion and lattice deformation during a galvanostatic insertion half cycle. The present findings contribute to a multiscale understanding of how lattice deformations, in addition to composition phase separation, affect microstructural evolution pathways. Specifically, in Li$_2$Mn$_2$O$_4$, we find that the tetragonal lattice variants nucleate independently in the electrode particle and form compatible twins during phase transformation. These twinned microstructures evolve---by adapting domain thickness and orientation---to lower the misfit strains at the phase boundaries. Our findings quantitatively estimate stress field concentrations in a typical Li$_2$Mn$_2$O$_4$ electrode during a discharge half cycle and suggest a possible mechanism for structural degradation in Li$_2$Mn$_2$O$_4$. More generally, our work establishes a theoretical framework that rigorously couples a Cahn-Hilliard type of diffusion with nonlinear gradient elasticity theory. This framework would be applicable to other symmetry-lowering first-order phase transformation materials beyond intercalation compounds.

\section*{Acknowledgments}
\noindent T. Zhang, D. Zhang, and A. Renuka Balakrishna acknowledge funding from the Air Force Fiscal Year 2023 Young Investigator Research Program (YIP), United States under Grant No. FOA-AFRL-AFOSR-2022-0005. The authors acknowledge the Center for Advanced Research Computing at the University of Southern California and the Center for Scientific Computing at University of California, Santa Barbara for providing resources that contributed to the research results reported in this paper. We thank Dr. Shiva Rudraraju (University of Wisconsin-Madison) for valuable discussions on the project. 

%Let's not share the link for the arxived version. 
%\section*{Code Availability}
%\noindent All codes developed are documented on \href{https://osf.io/gfnuh/?view_only=e1987cd883194117b56272401a23e9cf}{OSF|Solids \& Materials Group (UCSB)} and will be made available to public after the peer-review process.

\newpage
\begin{appendices}     
\appendix
\renewcommand{\thesection}{Appendix \Alph{section}} 
\section{Deriving the macroscopic force balance} \label{A}
\noindent We rewrite Eq.~(\ref{23}) using index notation:
\begin{eqnarray}\label{eq:A1}
0&=&\int_{\mathcal{P}}T_{R_{iJ}} \widetilde{\chi}_{i,J} ~dV+ \int _{\mathcal{P}} Y_{iJK}  \widetilde{\chi}_{i,JK} ~dV-\int_{\partial \mathcal{P}_i} \widetilde{\chi}_i t_i ~dA\nonumber\\
&&- \int_{\partial \mathcal{P}_i} m_i\widetilde{\chi}_{i,L}\hat{n}_L~dA - \int_{\zeta^{L_i}} \widetilde{\chi}_i l_i~dL -\int_{\mathcal{P}} \widetilde{\chi}_ib_i~dV.
\end{eqnarray}

\noindent Integrating Eq.~(\ref{eq:A1}) by parts yields
\begin{eqnarray}\label{eq:A2}
0&=&-\int _{\mathcal{P}}T_{R_{iJ,J}} \widetilde{\chi}_i ~dV + \int_{\partial \mathcal{P}} T_{R_{iJ}} \widetilde{\chi}_i   \hat{n}_J  ~dA  -\underbrace{\int _{\mathcal{P}} Y_{iJK,K}  \widetilde{\chi}_{i,J} ~dV}_\text{Integral A} + \int_{\partial \mathcal{P}}  Y_{iJK} \widetilde{\chi}_{i,J}  \hat{n}_K  ~dA \nonumber \\ 
&& - \int_{\partial \mathcal{P}_i} \widetilde{\chi}_i t_i ~dA - \int_{\partial \mathcal{P}_i} m_i\widetilde{\chi}_{i,L}\hat{n}_L~dA - \int_{\zeta^{L_i}} \widetilde{\chi}_i l_i~dL -\int_{\mathcal{P}} \widetilde{\chi}_ib_i~dV.
\end{eqnarray}

\noindent Applying integration by parts again in Eq.~(\ref{eq:A2}) but only to Integral A, and using normal gradient operator $\nabla^n$($\equiv (\hat{n}_K\partial_K$)), we obtain
\begin{eqnarray} \label{eq:A3}
0&=&-\int _{\mathcal{P}}T_{R_{iJ,J}} \widetilde{\chi}_i ~dV + \int_{\partial \mathcal{P}} T_{R_{iJ}} \widetilde{\chi}_i  \hat{n}_J  ~dA\nonumber \\ 
&&
+ \int _{\mathcal{P}} Y_{iJK,JK}  \widetilde{\chi}_i ~dV -  \underbrace{\int_{\partial \mathcal{P}}  Y_{iJK,K} \widetilde{\chi}_i \hat{n}_J  ~dA}_\text{Integral B}+ \underbrace{\int_{\partial \mathcal{P}}  Y_{iJK} \widetilde{\chi}_{i,J}  \hat{n}_K  ~dA}_\text{Integral C}\nonumber\\
&&- \int_{\partial \mathcal{P}_i} \widetilde{\chi}_i t_i ~dA - \int_{\partial \mathcal{P}_i} m_i(\nabla^n\widetilde{\chi}_i)~dA - \int_{\zeta^{L_i}} \widetilde{\chi}_i l_i~dL-\int_{\mathcal{P}} \widetilde{\chi}_ib_i~dV.
\end{eqnarray}

\noindent Expanding Integral B of Eq. (\ref{eq:A3}) using surface gradient operator $\nabla^s_K$($\equiv (\partial_K-\hat{n}_K\hat{n}_I\partial_I$)) yields 
\begin{eqnarray}\label{eq:A4}
 \int_{\partial \mathcal{P}}  Y_{iJK,K} \widetilde{\chi}_i  \hat{n}_J  ~dA  &&=
  \int_{\partial \mathcal{P}}  (Y_{iJK,L} \delta_{LK})  \widetilde{\chi}_i \hat{n}_J  ~dA  \nonumber \\ 
&& =  \int_{\partial \mathcal{P}}  ((\nabla^nY_{iJK})\hat{n}_L + \nabla^s_L Y_{iJK} )  \delta_{LK} \widetilde{\chi}_i  \hat{n}_J  ~dA \nonumber \\ 
&& =  \int_{\partial \mathcal{P}}  ((\nabla^nY_{iJK}) \hat{n}_K + \nabla^s_K Y_{iJK}) \widetilde{\chi}_i \hat{n}_J  ~dA.
\end{eqnarray}

\noindent Next, expanding Integral C of Eq.~(\ref{eq:A3}) we obtain
\begin{eqnarray}\label{eq:A5}
\int_{\partial \mathcal{P}}  Y_{iJK} \widetilde{\chi}_{i,J}  \hat{n}_K  ~dA  &&=\int_{\partial \mathcal{P}} ((\nabla^n \widetilde{\chi}_i)\hat{n}_J + \nabla^s_J \widetilde{\chi}_i) Y_{iJK}  \hat{n}_K  ~dA  \nonumber \\ 
&& = \int_{\partial \mathcal{P}} (\nabla^n \widetilde{\chi}_i) Y_{iJK} \hat{n}_J  \hat{n}_K  ~dA+\underbrace{\int_{\partial \mathcal{P}} (\nabla^s_J \widetilde{\chi}_i)Y_{iJK}\hat{n}_K}_\text{Integral D}~dA.
\end{eqnarray}

\noindent Integral D of Eq. (\ref{eq:A5}) yields
\begin{eqnarray}\label{eq:A6}
\int_{\partial \mathcal{P}}(\nabla^s_J\widetilde{\chi}_i)Y_{iJK}\hat{n}_K~dA &=& \underbrace{\int_{\partial \mathcal{P}} \nabla^s_J (\widetilde{\chi}_i Y_{iJK}\hat{n}_K)~dA}_\text{Integral E}-\underbrace{\int_{\partial \mathcal{P}}  \widetilde{\chi}_i \nabla^s_J (Y_{iJK}  \hat{n}_K)~dA}_\text{Integral F}.
\end{eqnarray}
in which Integral F is expanded as
\begin{eqnarray}\label{eq:A7}
\int_{\partial \mathcal{P}}  \widetilde{\chi}_i \nabla^s_J (Y_{iJK}  \hat{n}_K)~dA=\int_{\partial \mathcal{P}}  \widetilde{\chi}_i \nabla^s_J (Y_{iJK})\hat{n}_K~dA+\int_{\partial \mathcal{P}}  \widetilde{\chi}_i (\nabla^s_J\hat{n}_K)Y_{iJK}~dA.
\end{eqnarray}

\noindent Using the following integral identity \citep{toupin1962elastic}
\begin{eqnarray}\label{eq:A8} 
\int_{\partial \mathcal{P}} \nabla^s_I(f... \hat{n}_J)~dA &=& \int_{\partial \mathcal{P}} (h_{LL}\hat{n}_I \hat{n}_J-h_{IJ}) f... ~dA+\int_{\zeta^L} 	\left[\left[\hat{n}^{\mathit{\Gamma}}_I \hat{n}_J f...   \right] \right]~dL.
\end{eqnarray}
which holds for any smooth tensor field $f$... defined at points of a smooth surface $\mathit{\Gamma}=\partial \mathcal{P}$ with boundary curve $\zeta^L$, we expand Integral E as:

\begin{eqnarray}\label{eq:A9}
\int_{\partial \mathcal{P}} \nabla^s_J (\widetilde{\chi}_i Y_{iJK}  \hat{n}_K)  ~dA &=&  \int_{\partial \mathcal{P}} (h_{LL}\hat{n}_J \hat{n}_K-h_{JK}) \widetilde{\chi}_i Y_{iJK} ~dA+ \int_{\zeta^L} \left[\left[\hat{n}^{\mathit{\Gamma}}_J \hat{n}_K \widetilde{\chi}_i Y_{iJK}  \right] \right]~dL.
\end{eqnarray}

\noindent Here $h_{LL}=-\nabla^s_L\hat{n}_L$. $h_{IJ} = -\nabla^s_I\hat{n}_J= -\nabla^s_J\hat{n}_I$ are components of the second fundamental form of a smooth part of the boundary and the vector  $\hat{\mathbf{n}}^{\mathit{\Gamma}} = \hat{\mathbf{k}} \times  \hat{\mathbf{n}}$, where $\hat{\mathbf{k}}$ is the unit tangent to the curve $\zeta^L$. By combining Eqs.~(\ref{eq:A3}), (\ref{eq:A5}), (\ref{eq:A6}) and (\ref{eq:A9}), we arrive at the index form of macroscopic force balance in section \ref{sec:2.5.1}:
\begin{eqnarray}\label{eq:A10}
0=&-&\int _{\mathcal{P}} \widetilde{\chi}_i (T_{R_{iJ,J}}-Y_{iJK,JK}) ~dV + \int_{\partial \mathcal{P}}\widetilde{\chi}_i \biggl(T_{R_{iJ}}  \hat{n}_J-Y_{iJK,K}\hat{n}_J-\nabla^s_J (Y_{iJK}\hat{n}_K) \biggr.\nonumber \\ 
&+& \biggl.(h_{LL}\hat{n}_J \hat{n}_K-h_{JK}) Y_{iJK}\biggr)  ~dA  +  \int_{\partial \mathcal{P}} (\nabla^n \widetilde{\chi}_i) Y_{iJK} \hat{n}_J  \hat{n}_K~dA + \int_{\zeta^{L_i}} \widetilde{\chi}_i  \left[\left[\hat{n}^{\mathit{\Gamma}}_J \hat{n}_K Y_{iJK}  \right] \right]~dL\nonumber \\ 
&-&\int_{\partial \mathcal{P}_i} \widetilde{\chi}_i t_i ~dA - \int_{\partial \mathcal{P}_i} m_i(\nabla^n\widetilde{\chi}_i)~dA - \int_{\zeta^{L_i}} \widetilde{\chi}_i l_i~dL-\int_{\mathcal{P}} \widetilde{\chi}_ib_i~dV.
\end{eqnarray}

\noindent The resulting mechanical boundary conditions are given by:
\begin{eqnarray}
T_{R_{iJ}}  \hat{n}_J-Y_{iJK,K}\hat{n}_J-\nabla^s_J(Y_{iJK}\hat{n}_K)+(h_{LL}\hat{n}_J \hat{n}_K-h_{JK}) Y_{iJK}&=&t_i,\nonumber\\
Y_{iJK} \hat{n}_J  \hat{n}_K&=&m_i,\nonumber\\
\left[\left[\hat{n}^{\mathit{\Gamma}}_J \hat{n}_K Y_{iJK}  \right] \right]&=&l_i.
\end{eqnarray}
The local macroscopic force balance is given as:
\begin{eqnarray}\label{eq:A12}
    T_{R_{iJ,J}}-Y_{iJK,JK}+b_i=0.
\end{eqnarray}

\noindent Next, we write the tensor form of Eq.~(\ref{eq:A10}) in section \ref{sec:2.5.1} as:
\begin{eqnarray}
0&=&-\int _{\mathcal{P}} \widetilde{\boldsymbol{\chi}} \cdot (\nabla \cdot \mathbf{T}_{\mathrm{R}}^{\top}-\nabla \cdot(\nabla \cdot \mathbf{Y}^{\top})^{\top}+\mathbf{b}) ~dV \nonumber \\ 
&&+ \int_{\partial \mathcal{P}} \widetilde{\boldsymbol{\chi}} \cdot \left(\mathbf{T}_{\mathrm{R}}\hat{\mathbf{n}}-(\nabla \cdot \mathbf{Y}^{\top})\hat{\mathbf{n}}-\nabla^s \cdot (\mathbf{Y} \cdot \hat{\mathbf{n}})^{\top}- \mathbf{Y}\colon ((\nabla^s \cdot \hat{\mathbf{n}})\hat{\mathbf{n}} \otimes \hat{\mathbf{n}}-\nabla^s \hat{\mathbf{n}})-\mathbf{t}\right)  ~dA \nonumber \\ 
&&+\int_{\partial \mathcal{P}} (\nabla\widetilde {\boldsymbol{\chi}})\hat{\mathbf{n}} \cdot (\mathbf{Y}\colon(\hat{\mathbf{n}} \otimes \hat{\mathbf{n}})-\mathbf{m})~dA 
+ \int_{\zeta^L} \widetilde{\chi}_i (\left[\left[\hat{n}^{\mathit{\Gamma}}_J \hat{n}_K Y_{iJK}  \right] \right]-l_i)~dL.
\end{eqnarray}
The mechanical boundary conditions are given by:
\begin{eqnarray}
\mathbf{T}_{\mathrm{R}}\hat{\mathbf{n}}-(\nabla \cdot \mathbf{Y}^{\top})\hat{\mathbf{n}}-\nabla^s \cdot (\mathbf{Y} \cdot \hat{\mathbf{n}})^{\top}-\mathbf{Y}\colon ((\nabla^s \cdot \hat{\mathbf{n}})\hat{\mathbf{n}} \otimes \hat{\mathbf{n}}-\nabla^s \hat{\mathbf{n}})&=&\mathbf{t},\nonumber\\
\mathbf{Y}\colon(\hat{\mathbf{n}} \otimes \hat{\mathbf{n}})&=&\mathbf{m},\nonumber\\
\left[\left[\hat{n}^{\mathit{\Gamma}}_J \hat{n}_K Y_{iJK}  \right] \right]&=&l_i,
\end{eqnarray}
and the local macroscopic force balance is:
\begin{eqnarray}\label{eq:A15}
  \nabla \cdot \mathbf{T}_{\mathrm{R}}^{\top}-\nabla \cdot(\nabla \cdot \mathbf{Y}^{\top})^{\top}+\mathbf{b}=0.
\end{eqnarray}

\noindent Recall from Eq.~(\ref{51}) of the main text, the symmetry condition for the third-order stress tensor $\mathbf{Y}$ can be written as $Y_{iJK} = Y_{iKJ}$. By combining the symmetry constraint of $\mathbf{Y}$ with Eqs.~(\ref{eq:A4}), (\ref{eq:A7}), and (\ref{eq:A10}), we derive the final index notation for the macroscopic force balance as follows:
\begin{eqnarray}\label{eq:A16}
0=&-&\int _{\mathcal{P}} \widetilde{\chi}_i (T_{R_{iJ,J}}-Y_{iJK,JK}+b_i) ~dV + \int_{\partial \mathcal{P}} \widetilde{\chi}_i \biggl(T_{R_{iJ}}  \hat{n}_J-(\nabla^n Y_{iJK})\hat{n}_J\hat{n}_K-2\nabla^s_J (Y_{iJK})\hat{n}_K\biggr.\nonumber \\ 
&-&Y_{iJK}\nabla^s_J\hat{n}_K+\biggl.(h_{LL}\hat{n}_J \hat{n}_K-h_{JK}) Y_{iJK}\biggr)  ~dA  +  \int_{\partial \mathcal{P}} (\nabla^n\widetilde{\chi}_i) Y_{iJK} \hat{n}_J \hat{n}_K  ~dA\nonumber \\ 
&+&\int_{\zeta^{L_i}} \widetilde{\chi}_i  \left[\left[\hat{n}^{\mathit{\Gamma}}_J \hat{n}_K Y_{iJK}  \right] \right]~dL-\int_{\partial \mathcal{P}_i} \widetilde{\chi}_i t_i ~dA - \int_{\partial \mathcal{P}_i} m_i(\nabla^n\widetilde{\chi}_i)~dA - \int_{\zeta^{L_i}} \widetilde{\chi}_i l_i~dL.
\end{eqnarray}

\noindent The index notation representing the local macroscopic force balance remains consistent with Eq.~(\ref{eq:A12}). However, the mechanical boundary conditions are derived in the following manner:
\begin{eqnarray}
T_{R_{iJ}}  \hat{n}_J-(\nabla^n Y_{iJK})\hat{n}_J\hat{n}_K-2\nabla^s_J (Y_{iJK})\hat{n}_K-Y_{iJK}\nabla^s_J\hat{n}_K&&\nonumber\\
+(h_{LL}\hat{n}_J \hat{n}_K-h_{JK}) Y_{iJK}&=&t_i,\nonumber\\
Y_{iJK} \hat{n}_J  \hat{n}_K&=&m_i,\nonumber\\
\left[\left[\hat{n}^{\mathit{\Gamma}}_J \hat{n}_K Y_{iJK}  \right] \right]&=&l_i.
\end{eqnarray}

\noindent The tensor form of the macroscopic force balance in Eq.~(\ref{eq:A16}) is expressed as:
\begin{eqnarray}
0&=&-\int _{\mathcal{P}} \widetilde{\boldsymbol{\chi}} \cdot (\nabla \cdot \mathbf{T}_{\mathrm{R}}^{\top}-\nabla \cdot(\nabla \cdot \mathbf{Y}^{\top})^{\top}+\mathbf{b}) ~dV \nonumber \\ 
&&+ \int_{\partial \mathcal{P}} \widetilde{\boldsymbol{\chi}} \cdot \biggl(\mathbf{T}_{\mathrm{R}}\hat{\mathbf{n}}-(\nabla\mathbf{Y}\cdot\hat{\mathbf{n}})\colon(\hat{\mathbf{n}} \otimes \hat{\mathbf{n}})-2(\nabla^s\cdot (\mathbf{Y}^{\top})^{\top})^{\top}\hat{\mathbf{n}}\nonumber \biggr.\\
&&\qquad\biggl.-\mathbf{Y}\colon\nabla^s\hat{\mathbf{n}}- \mathbf{Y}\colon ((\nabla^s \cdot \hat{\mathbf{n}})\hat{\mathbf{n}} \otimes \hat{\mathbf{n}}-\nabla^s \hat{\mathbf{n}})-\mathbf{t}\biggr)  ~dA \nonumber \\ 
&& +  \int_{\partial \mathcal{P}} (\nabla\widetilde {\boldsymbol{\chi}})\hat{\mathbf{n}} \cdot (\mathbf{Y}\colon(\hat{\mathbf{n}} \otimes \hat{\mathbf{n}})-\mathbf{m})~dA 
+ \int_{\zeta^L} \widetilde{\chi}_i (\left[\left[\hat{n}^{\mathit{\Gamma}}_J \hat{n}_K Y_{iJK}  \right] \right]-l_i)~dL.
\end{eqnarray}
The tensor form of the local macroscopic force balance remains unchanged from Eq.~(\ref{eq:A15}), whereas parts of mechanical boundary conditions in section \ref{sec:3.4} are described as follows:
\begin{eqnarray}
    \mathbf{T}_{\mathrm{R}}\hat{\mathbf{n}}-(\nabla\mathbf{Y}\cdot\hat{\mathbf{n}})\colon(\hat{\mathbf{n}} \otimes \hat{\mathbf{n}})-2(\nabla^s\cdot (\mathbf{Y}^{\top})^{\top})^{\top}\hat{\mathbf{n}}-\mathbf{Y}\colon\nabla^s\hat{\mathbf{n}}&&\nonumber\\
    - \mathbf{Y}\colon ((\nabla^s \cdot \hat{\mathbf{n}})\hat{\mathbf{n}} \otimes \hat{\mathbf{n}}-\nabla^s \hat{\mathbf{n}}) &=& \mathbf{t},\nonumber\\
    \mathbf{Y}\colon(\hat{\mathbf{n}} \otimes \hat{\mathbf{n}}) &=&\mathbf{m},\nonumber\\
    \left[\left[\hat{n}^{\mathit{\Gamma}}_J \hat{n}_K Y_{iJK}  \right] \right]&=&l_i.
\end{eqnarray}

\newpage
\section{Implementing the galvanostatic (dis-)charge condition}\label{Appendix B}
\noindent Using the relationships $\mathbf{j}=\frac{D_0c_0}{L}\bar{\mathbf{j}}$ and $I=k_0L^2\bar{I}$ we write the galvanostatic condition in Eq.~(\ref{eq69}) as follows: 

\begin{align}
    \sum_k\int_{\partial \bar{\Omega}^{\{\mathbf{j}\}_k}}\bar{j_n} \ d\bar{A} = \mathrm{Da} \times \bar{I}.
    \label{eq:totalReactionrate_DLess}
\end{align}

\noindent In Eq.~(\ref{eq:totalReactionrate_DLess}) the dimensionless global flux $\bar{I}$ is applied on the corresponding dimensionless reactive boundaries $\partial \bar{\Omega}^{\{\mathbf{j}\}}$. The dimensionless flux is given by:
\begin{align}
    \bar{j_n}=\mathrm{Da}(1-\bar{c})[\mathrm{exp}(-0.5\Delta \bar{\phi})-\mathrm{exp}(\bar{\mu})\mathrm{exp}(0.5\Delta \bar{\phi})],
    \label{eq:Jn}
\end{align}

\noindent in which, $\Delta \bar{\phi}=\frac{F \Delta \phi}{RT_0}$.
Substituting Eq.~(\ref{eq:Jn}) into Eq.~(\ref{eq:totalReactionrate_DLess}) and accounting for $m$ active surface areas, we obtain
\begin{align}
    z\sum_{k=1}^m\int_{\partial \bar{\Omega}^{\{\mathbf{j}\}_k}}(1-\bar{c})d\bar{A}-\frac{1}{z}\sum_{k=1}^m\int_{\partial \bar{\Omega}^{\{\mathbf{j}\}_k}}(1-\bar{c})\mathrm{exp}(\bar{\mu})d\bar{A}-\bar{I}=0,\nonumber\\
    z^2\sum_{k=1}^m\int_{\partial \bar{\Omega}^{\{\mathbf{j}\}_k}}(1-\bar{c})d\bar{A}-\sum_{k=1}^m\int_{\partial \bar{\Omega}^{\{\mathbf{j}\}_k}}(1-\bar{c})\mathrm{exp}(\bar{\mu})d\bar{A}-\bar{I}z=0.
    \label{eq:y}
\end{align}

\noindent Here, $z=\mathrm{exp}(-0.5\Delta \bar{\phi})$ and we relabel the integrals as:
\begin{align}
    \begin{cases}
    \mathrm{int}^-=\sum_{k=1}^m\int_{\partial \bar{\Omega}^{\{\mathbf{j}\}_k}}(1-\bar{c})d\bar{A}\\
    \mathrm{int}^+=\sum_{k=1}^m\int_{\partial \bar{\Omega}^{\{\mathbf{j}\}_k}}(1-\bar{c})\mathrm{exp}(\bar{\mu})d\bar{A}\\
    \end{cases}
    \label{eq:intSB}
\end{align}

\noindent By combining Eq.~(\ref{eq:y}) and Eq.~(\ref{eq:intSB}), we obtain
\begin{align}
    \mathrm{int^-}z^2-\bar{I}z-\mathrm{int^+}=0,\nonumber\\
    z=\frac{1}{2\mathrm{int^-}}\left[\bar{I}\pm\sqrt{(\bar{I})^2+4\mathrm{int^-int^+}}\right].
\end{align}

\noindent Here, we consider $z$ as a positive solution in order to compute the voltage drop $\Delta \bar{\phi} = -2\mathrm{ln}z$. Please note that the value of $\Delta \bar{\phi}$ is computed and substituted into the Bulter-Volmer equation at every time step in our calculations.

\section{Determining the multiwell potential $\bar{\psi}_{\mathrm{ther}}$} \label{C}
\noindent First, the chemical potential $\mu$ is related to the open circuit voltage $E_{\mathrm{oc}}$ by 
\begin{eqnarray} \label{eq13} 
E_{\mathrm{oc}}\left(\bar{c},T\right)=-\frac{1}{e N_{A}}\mu\left(\bar{c},T\right).
\end{eqnarray}
Following \citep{zhang2021microstructure} we write the chemical potential $\mu$ as 
\begin{eqnarray} \label{eq14} 
\mu\left(\bar c,T\right)=\left\{ 
\begin{array}{cc}
RT_0\left(\frac{\overline \psi_{\mathrm{ther}}\left(\overline {c}_2,T\right)- \overline{ \psi}_{\mathrm{ther}}\left(\bar{c}_1,T\right)}{\bar{c}_2-\bar{c}_1}\right) & \mbox{if} \ \  \bar{c}_1 \le \bar{c} \le \bar{c}_2\\
RT_0\frac{\partial \bar \psi_{\mathrm{ther}}}{\partial \bar{c}} & \mbox{for}\ \  \mbox{otherwise}
\end{array}\right.,
\end{eqnarray}
in which $\bar{c}_1$ and $\bar{c}_2$ are the binodal concentrations that are found by constructing a common tangent to the multiwell potential curve (Maxwell construction). 

\vspace{2mm}
\noindent Using Eqs. (\ref{eq13}) and (\ref{eq14}), we fit the OCV to the experimental data \citep{thackeray1983lithium} and derive the unknown parameters.
We obtain a good fit with the experimental OCV with $n = 3, \overline{\mu}_0=-579.454, \alpha_1=-926.715, \alpha_2=-926.715$, and $\alpha_3=-470.114$. This ensures that phase segregation occurs at the two binodal concentrations $\bar{c}_1=0.501$ and $\bar{c}_2=0.99$ 
with the Maxwell construction given by 
\begin{eqnarray} 
\frac{\partial \bar{\psi}_{\mathrm{ther}}(\bar{c}_1)}{\partial \bar{c}} =
\frac{\partial \bar{\psi}_{\mathrm{ther}}(\bar{c}_2)}{\partial \bar{c}} =
\frac{\bar{\psi}_{\mathrm{ther}}\left(\bar{c}_2\right)-\bar{\psi}_{\mathrm{ther}}\left(\bar{c}_1\right)}{\bar{c}_2-\bar{c}_1}.
\end{eqnarray}

% \section{\textcolor{red}{Numerical computation details}}
% \begin{itemize}
%     \item The particle domain is discretized on a $50 \times 50$ mesh.
%     \item In order to improve the convergence, we use the equation describing the mass balance (Eq.~(\ref{eq48})) to solve for the chemical potential rather than the species concentration.
% \end{itemize}

\section{Symbols} \label{D}
\noindent We summarize all symbols used in our work in Table~\ref{T2} below.

\begin{longtable}{p{2.4cm}p{8.5cm}p{2.4cm}}
  \caption{Summary of symbols.} \\ 
  \label{T2}\\
  \hline
  Symbol&Description&Unit\\
    \hline
    $\Omega$&The reference body&[/]\\
    $\partial \Omega$&Surface of the reference body&[/]\\
    $\mathcal{P}$&Arbitrary part of the reference body&[/]\\
    $\partial \mathcal{P}$&Surface of arbitrary part of the reference body&[/]\\
    $\zeta^L$&Smooth boundary edge &[/]\\
    $c$&Species concentration in the reference configuration&[$\mathrm{mol/m^3}$]\\
    $c_0$&Maximum reference species concentration&[$\mathrm{mol/m^3}$]\\
    $L$&Characteristic length&[$\mathrm{m}$]\\
    $t$&Time&[$\mathrm{s}$]\\    
    $D_0$&Diffusion coefficient&[$\mathrm{m^2/s}$]\\
    % $k_B$&Boltzman constant&[$\mathrm{J/K}$]\\
    $R$&Gas constant&[$\mathrm{J/(mol\cdot K})$]\\
    $N_A$&Avogadro constant&[$\mathrm{1/mol}$]\\
    $T_0$&Reference temperature&[$\mathrm{K}$]\\
    $T$&Temperature&[$\mathrm{K}$]\\
    $C_{11}/C_{12}/C_{44}$&Elastic constants&[Pa]\\
    $\beta_1$&Deviatoric modulus&[Pa]\\
    $K$&Bulk modulus&[Pa]\\
    $G$&Shear modulus&[Pa] \\
   $\beta_2/\beta_3$&Nonlinear elastic constants&[Pa]  \\
   $\Delta V$&Volume change&[/]\\
   $f$&Volume fraction&[/]\\
   $i$&Current density&[$\mathrm{A/m^2}$]\\
   $i_0$&Exchange current density&[$\mathrm{A/m^2}$]\\
   $k_0$&Reaction rate constant&[$\mathrm{mol/(m^2\cdot s)}$]\\
   $\beta$&Electron-transfer symmetry factor&[/]\\
   $F$&Faraday constant&[$\mathrm{C/mol}$]\\
   $I$&Global flux&[$\mathrm{mol/s}$]\\
   $\mathcal{D}$&Dissipation density &[$\mathrm{J/(m^3 \cdot s)}$]\\
   $\alpha_{i}$&Coefficients representing the weight of enthalpy &[/]\\
   $\mu_+$&Chemical potential in the electrolyte&[$\mathrm{J/mol}$]\\
   $\mu_0$&Reference chemical potential&[$\mathrm{J/mol}$]\\
    $\mu$&Chemical potential in the reference configuration&[$\mathrm{J/mol}$]\\
    $\eta$&Surface overpotential&[V]\\
    $Da$&Damköhler number&[/]\\
    $\Delta \phi$&Voltage drop&[V]\\
   $\psi_{\mathrm{bulk}}$&Bulk free energy density&[$\mathrm{J/m^3}$]\\
   $\psi_{\mathrm{grad}}$&Gradient energy density&[$\mathrm{J/m^3}$]\\
   % $\psi_{\mathrm{ther}}$&Thermodynamic free energy density&[$\mathrm{J/m^3}$]\\
   % $\psi_{\mathrm{elas}}$&Elastic energy density&[$\mathrm{J/m^3}$]\\
   % $\psi_{\mathrm{coup}}$&Coupled energy density&[$\mathrm{J/m^3}$]\\
   $\lambda$&Concentration gradient energy coefficient&[$\mathrm{m^2}$]\\ 
  $\kappa$&Strain gradient energy coefficient&[$\mathrm{m^2}$]\\
  $W_{\mathrm{ext}}$&External power&$[\mathrm{J/s}]$\\
  $W_{\mathrm{int}}$&Internal power power&$[\mathrm{J/s}]$\\
  $e_i$&Strain measures&[/]\\
  % $e_2^m$&Strain minimum &[/]\\
  $E_{oc}$&Open circuit voltage &[V]\\
  $\nabla$ & Gradient operator&[$1/m$]\\
  $\nabla^n$&Normal gradient operator &[1/m]\\
  $\nabla^s$&Surface gradient operator &[1/m]\\
 $\mathbf{x}$ &Material points&[$\mathrm{m}$]\\
 % $\mathbf{x}$ &Spatial points&[$m$]\\
 $\boldsymbol{\chi}$ &Mapping from material to spatial frame&[$\mathrm{m}$]\\
 $\boldsymbol{\xi}$ &Vector microscopic force&[$\mathrm{N/m}$]\\
  $\mathbf{u}$&Displacement&[$\mathrm{m}$]\\
  % $\mathbf N$&Surface normal in the reference configuration&[/]\\
  % $\mathbf n$&Surface normal in the current configuration&[/]\\
   $\mathbf{j}$&Species flux in the reference configuration&[$\mathrm{mol/(m^2\cdot s})$]\\
  $\mathbf{t}$&Surface traction &[Pa]\\
  $\mathbf{m}$&Surface moment &[$\mathrm{N/m}$]\\
  $\mathbf{l}$&Line force &[$\mathrm{N/m}$]\\
  $\mathbf{b}$&Body force &[$\mathrm{N/m^3}$]\\
  $\mathbf{\hat{n}}/ \mathbf{\hat{m}}/ \mathbf{a}/ \mathbf{b}$&Vector&[/]\\
  $\mathbf{e}_i$&Lattice vector &[/]\\
  $\mathbf{U}_i/\mathbf{U}_j$& Deformation tensor&[/]\\
  $\mathbf{Q}/\mathbf{Q}'$&Rotation tensor&[/]\\
  $\mathbf{K}$&Twin plane direction&[/]\\
  $\zeta$&Scalar microscopic traction&[$\mathrm{N/m}$]\\
  $\pi$&Scalar microscopic force &[$\mathrm{N/m^2}$]\\
  $\mathbf{I}$($\delta_{{i J}}$)&Second order unit tensor&[/]\\
  $\mathbf{F}$&Deformation gradient&[/]\\
  $\mathbf{\mathbf{T}_R}$&First Piola-Kirchhoff stress tensor&[Pa]\\
  $\mathbf{Y}$&Third-order stress tensor&[$\mathrm{Pa\cdot m}$]\\
  $\mathbf{E^{0}}$&Spontaneous strain&[/]\\
 $\boldsymbol{\rho}$&Lagrange multipliers &[Pa]\\
  $\mathbf{M}$&Mobility tensor &[$\mathrm{mol^2/(m\cdot J\cdot s)}$]\\
  % $\boldsymbol{\varepsilon^{0}}$&Stress-free strain  &[/]\\
    \hline
   % \end{tabular*}
%\end{table*}
\end{longtable}
\end{appendices}

\newpage
\bibliographystyle{elsarticle-harv} 
\bibliography{LMO}

\end{document}